\documentclass[onecolumn, fleqn,usenatbib]{mnras}

\usepackage{newtxtext,newtxmath}
\usepackage[T1]{fontenc}
\usepackage{lastpage} % Resolve the final page after all floats are placed.
\def\apjl{ApJ}%% Astrophysical Journal, Letters
\def\apjs{ApJS}%% Astrophysical Journal, Supplement
\def\ao{Appl.~Opt.}%% Applied Optics
\def\aap{A\&A}%% Astronomy and Astrophysic
\newcommand{\be}{\begin{equation}}
\newcommand{\ee}{\end{equation}}
\newcommand{\bary}{\begin{eqnarray}}
\newcommand{\eary}{\end{eqnarray}}

\DeclareRobustCommand{\VAN}[3]{#2}
\let\VANthebibliography\thebibliography
\def\thebibliography{\DeclareRobustCommand{\VAN}[3]{##3}\VANthebibliography}

\usepackage{epstopdf}
\usepackage{ulem}

\usepackage{amssymb}
\usepackage{times}
\usepackage{graphicx}
\usepackage[usenames,dvipsnames]{pstricks}
\usepackage{psfrag}
\usepackage{lipsum}
\usepackage{natbib}
\usepackage{textcase}
\usepackage{orcidlink}

\usepackage{multirow}
\usepackage{graphicx}

\usepackage[lofdepth,lotdepth]{subfig} %%%%Package added by 
\title[Early X-ray emission of GRB 241030A]
{GRB 241030A: A fallback-accreting magnetar description of the early X-ray emission}

\author[Fraija, N. et al.]{
Nissim Fraija\orcidlink{0000-0002-0173-6453}$^{1}$\thanks{E-mail: nifraija@astro.unam.mx}, Simone Dichiara\orcidlink{0000-0001-6849-1270}$^{2}$ and Antonio Galv\'an\orcidlink{0000-0001-5193-3693}$^{1}$\\
$^{1}$Instituto de Astronom\' ia, Universidad Nacional Aut\'onoma de M\'exico, Circuito Exterior, C.U., A. Postal 70-264, 04510 M\'exico City, M\'exico\\
$^{2}$ Department of Astronomy and Astrophysics, The Pennsylvania State University, 525 Davey Lab, University Park, PA 16802, USA
}

\date{Accepted XXX. Received YYY; in original form ZZZ}

\pubyear{\the\year{}}

\begin{document}
\label{firstpage}
\pagerange{\pageref{firstpage}--\pageref{LastPage}}
\maketitle

% Abstract of the paper
\begin{abstract}
{ The bright, variable early X-ray emission of GRB~241030A at $z=1.411$
motivates investigating continued central-engine activity before the
smoother afterglow emerges. We model a possible underlying magnetar
contribution to the unabsorbed 0.3--10~keV XRT flux over 80--430~s,
evolving the spin under fallback-disk and electromagnetic torques.
A descriptive estimate gives $P_0\simeq4.25$~ms,
$B\simeq3.04\times10^{15}$~G and source-frame $t_{\rm fb}\simeq143$~s,
with exploratory temporal-block uncertainty ranges. The effective
efficiency reaches its geometry-dependent ceiling of 81.2.
At observer time 200~s, this realization has $P_*\simeq4.82$~ms and
magnetospheric, corotation and light-cylinder radii
$r_{\rm m}\simeq235$~km, $r_{\rm co}\simeq48$~km and
$r_{\rm LC}\simeq230$~km. It is on the ejector branch; light-cylinder
entry near 255~s initiates negative propeller torque and steepens the
decline. The smooth component declines throughout the fitted interval;
additional variable emission contributes to the observed maximum near
180~s. A correlated-residual Bayesian analysis permits a fainter
component and broader dynamical histories, with median $P_0=16.8$~ms
and central 68 per cent interval 2.76--286~ms. However, posterior
predictive checks reveal that this statistical description typically
underpredicts the observed flux level and variability, limiting its
physical interpretation. Fallback-modified magnetar activity remains
a possible descriptive scenario, but these analyses do not establish
a unique underlying component or robust constraints on the initial
spin and disk-interaction history. We also discuss Lense--Thirring
precession and alternative mechanisms as possible origins of the
early X-ray variability.}
\end{abstract}

\begin{keywords}
gamma-ray burst: individual: GRB 241030A --
stars: magnetars --
accretion, accretion discs --
relativistic processes --
X-rays: bursts
\end{keywords}

%%%%%%%%%%%%%%%%%%%%%%%%%%%%%%%%%%%%%%%%%%%%%%%%%%

%%%%%%%%%%%%%%%%% BODY OF PAPER %%%%%%%%%%%%%%%%%%

\section{Introduction}
\label{sec:introduction}

{ Gamma-ray bursts (GRBs) are powered by ultrarelativistic outflows launched by newly formed compact objects. A stellar-mass black hole (BH) surrounded by a hyperaccreting disk is commonly invoked as the central engine of long-duration GRBs, although a rapidly rotating and strongly magnetized neutron star (NS) represents another viable possibility \citep[e.g.][]{1992Natur.357..472U,1998MNRAS.298...87D,2001ApJ...559..110Z, 2006MNRAS.372L..19F, Troja2007, Lyons2010, 2014ApJ...785...74L}

In the latter scenario, the extraction of rotational energy from a millisecond magnetar may sustain prolonged high-energy activity. Early X-ray emission is particularly useful for investigating this possibility because flares, plateaus, and other structured components may retain information about continuing central-engine activity that is difficult to infer from the smoother late-time afterglow. }

The evolution of a newborn magnetar can be substantially modified by fallback accretion. Material that remains gravitationally bound after the collapse may return to the compact remnant and form an accretion disk. The interaction between this disk and the stellar magnetosphere changes the torque, spin evolution, and fraction of open magnetic flux, thereby modifying the electromagnetic luminosity relative to that of an isolated magnetar \citep{2018ApJ...857...95M}. Depending on the relative locations of the magnetospheric, corotation, and light-cylinder radii, the system may evolve between propeller and accretion regimes. Models combining fallback accretion and magnetic spin-down have consequently been considered for extended emission, X-ray plateaus, and late flares in GRBs \citep[e.g.][]{2014MNRAS.438..240G,2017MNRAS.470.4925G, 2021ApJ...918...12F}. Recent applications of this framework to the kilonova-associated GRB~211211A and the short GRB~250704B/EP250704a have further shown that fallback-modified magnetar activity can sustain prolonged high-energy or X-ray emission and continued energy injection \citep{2026MNRAS.549ag949F, 2026MNRAS.550g1295F}.

{ Rapid GRB variability can arise from several emission and dynamical processes. Its interpretation requires separating the secular component from correlated and non-stationary structure; timing searches in GRB~230307A illustrate this difficulty \citep{Huppenkothen2025}. Here we treat the variable excesses statistically and restrict the physical inference to a possible smooth magnetar contribution.}

GRB 241030A was detected on 2024 October 30 by the Fermi/Gamma-ray Burst Monitor (GBM), the Fermi Large Area Telescope (LAT), and the Swift/Burst Alert Telescope \citep[BAT;][]{FermiGBMTeam2024, Pillera2024, Klingler2024}. It is a long-duration GRB at a redshift of $z=1.411$ \citep{deBarra2024,Zheng2024}. Swift X-Ray Telescope (XRT) began observing the source approximately 80 s after the trigger and detected a bright, strongly variable X-ray episode extending to approximately 430 s, with a maximum near 180 s. Previous studies have examined its prompt thermal component, rapidly variable optical emission, and broadband afterglow in the context of photospheric emission, internal and reverse shocks, and forward-shock models \citep{Wang2025,Varun2026,Ducoin2026}.  The broadband afterglow analysis of \cite{Ducoin2026} uses observations obtained after approximately 400 s. Their structured-jet models reproduce the late-time observations but require large kinetic energies, small microphysical parameters, and a prompt radiative efficiency below $10^{-3}$, leading the authors to question whether the afterglow is produced exclusively by standard forward-shock emission. The X-ray light curve remains structured until approximately 430 s and subsequently approaches a smoother power-law decay. We therefore regard the interval between approximately 400 and 430 s as a possible transition or overlap between the structured early emission and the later broadband afterglow.

{ In this work, we investigate a possible smooth magnetar contribution to
the unabsorbed 0.3--10 keV XRT flux over the observer-frame interval
$80\ {\rm s}\lesssim t_{\rm obs}-t_0\lesssim430\ {\rm s}$, where
$t_0$ is the Swift/BAT trigger time. The dynamics are evaluated at the
corresponding source-frame epochs. We distinguish inference under a
temporally correlated residual model from a separate marginal descriptive
estimate with temporal-block uncertainty. Joint parameter samples are
propagated into the stellar spin and characteristic radii. Neither a
periodic signal nor a timing-based radius hierarchy is required by this
analysis. 
Section~\ref{sec:data} describes the observations and data selection.
Section~\ref{sec:model} specifies the fallback-magnetar and statistical
models. Section~\ref{sec:results} presents the flux descriptions and
jointly propagated dynamics. Sections~\ref{sec:discusion} and
\ref{sec:conclusions} discuss the interpretation and conclusions.}

{ \section{X-ray observations and data selection}}
\label{sec:data}

\subsection{Swift observations and data selection}
\label{sec:swift_data}
The Swift/BAT triggered on GRB~241030A at $t_0=05{:}48{:}03$ UT on 2024 October 30, which we adopt as the temporal reference throughout this work \citep{Klingler2024}. The Fermi/GBM triggered essentially simultaneously, at 05:48:03.33 UT \citep{deBarra2024}. BAT measured a duration of $T_{90}=173.3\pm5.0$ s in the 15--350 keV energy band \citep{Klingler2024,Krimm2024}. The Swift/XRT began observing the field approximately 73.5 s after the BAT trigger, and the post-slew settling was completed at $t_{\rm obs}-t_0\simeq79.8$ s \citep{Klingler2026}.  Optical spectroscopy obtained with the Low Resolution Imaging Spectrometer (LRIS) on the Keck I telescope established a redshift of $z=1.411$ \citep{Zheng2024}.

We use the publicly available unabsorbed 0.3--10 keV XRT flux light curve obtained from the UK Swift Science Data Centre \citep{Evans2007,Evans2009}. The early X-ray emission is characterized by a bright and strongly variable episode extending from approximately 80 to 430 s after $t_0$, with a maximum near 180 s. After approximately 430 s, the XRT light curve approaches a smoother power-law decay \citep{Wang2025,Ducoin2026,Klingler2026}.

{ The broadband afterglow analysis of \citet{Ducoin2026} uses observations obtained after $t_{\rm obs}-t_0=400$ s. Because the bright variable X-ray episode continues until approximately 430 s, we regard the interval between approximately 400 and 430 s as a possible transition or overlap between the structured early emission and the subsequent afterglow. We therefore restrict the fallback-accreting magnetar fit to $80~{\rm s}\lesssim t_{\rm obs}-t_0\lesssim430~{\rm s}$.  The model is not intended to describe the subsequent broadband evolution. Later X-ray measurements and the BAT light curve are shown only to provide temporal context. Only the unabsorbed XRT flux measurements enter the magnetar likelihood.}

\subsection{X-ray luminosity}

We use the unabsorbed observer-frame 0.3--10 keV XRT flux and compare it
directly with the bin-averaged magnetar prediction. At $z=1.411$, this
band corresponds to 0.7233--24.11 keV in the source frame. The effective
efficiency is defined in that matched band, so no additional
photon-index-based $K$-correction is assigned to the magnetar component.
{ We adopt $D_L=10.30$ Gpc and evaluate the dynamics at
$t=(t_{\rm obs}-t_0)/(1+z)$. The figures retain observer-frame time.
The current analysis uses 380 original bins with centres in 80--430 s.}

%The unabsorbed XRT flux was converted into an isotropic-equivalent X-ray luminosity according to
%\begin{equation}
%L_{\rm X,iso}(t_{\rm obs})=4\pi D_L^2 F_X(t_{\rm obs})K_X,
%\label{eq:Lxobs}
%\end{equation}
%where $F_X$ is the unabsorbed flux measured in the 0.3--10 keV observer-frame band and $K_X$ is the corresponding K-correction. At $z=1.411$, we adopt a luminosity distance of $D_L=10.30$ Gpc, consistent with a standard flat $\Lambda$CDM cosmology \citep{2020A&A...641A...6P}.

%For a power-law photon spectrum, $N(E)\propto E^{-\Gamma_X}$, the correction from the observed 0.3--10 keV band to the same numerical energy interval in the source frame is $K_X=(1+z)^{\Gamma_X-2}$. For each XRT measurement, $K_X$ is calculated using the photon index associated with its corresponding spectral interval, thereby accounting approximately for the spectral evolution during the bright X-ray episode. The luminosity measurements are compared with the magnetar model at the corresponding source-frame epochs, $t=\frac{t_{\rm obs}-t_0}{1+z}$. Thus, the XRT times supplied to the luminosity fit are divided by $(1+z)$, whereas model curves displayed against observer-frame time are evaluated using $t_{\rm obs}-t_0=(1+z)t$. Unless explicitly stated otherwise, observational times are quoted in the observer frame, while the dynamical quantities introduced in Section~\ref{sec:model} are calculated in the source frame.

{ \subsection{Scope of the timing interpretation}
\label{sec:timing-scope}
The present analysis addresses the smooth X-ray contribution and its
conditional dynamical implications. We do not undertake a new periodicity
search, assume a periodic signal, or use a reported timing centroid as a
constraint. No timing-based precession radius, outer precessing boundary,
or global precession-period evolution is inferred. The illustrative
local timescales in Section~\ref{sec:discussion_lt} are calculated under
additional assumptions and do not enter the inference. The statistical excess component
represents variability without identifying its physical mechanism.}

\section{Fallback-accreting magnetar model}
\label{sec:model}

We consider a newly born magnetar surrounded by a fallback disk. The disk modifies the stellar spin evolution through the exchange of angular momentum and can enhance the electromagnetic spin-down power by opening magnetic field lines that would remain closed in an isolated magnetosphere \citep{2011ApJ...736..108P,2016ApJ...822...33P,2018ApJ...857...95M}. We prescribe the fallback accretion rate and evolve the disk--magnetosphere interaction and stellar angular momentum self-consistently.  All dynamical quantities in this section are evaluated in the source frame.

\subsection{Fallback accretion rate}
\label{sec:fallback}

The return rate of gravitationally bound material is described phenomenologically by the smooth broken-power-law prescription

\begin{equation}
\dot{M}_{\rm fb}(t)
=
\dot{M}_0
\left[
\left(\frac{t}{t_{\rm fb}}\right)^{-1/2}
+
\left(\frac{t}{t_{\rm fb}}\right)^{5/3}
\right]^{-1},
\label{eq:fallback_rate}
\end{equation}

where $t$ is measured in the source frame. This prescription approaches $\dot{M}_{\rm fb}\propto t^{1/2}$ at early times and the canonical $\dot{M}_{\rm fb}\propto t^{-5/3}$ behaviour at late times \citep{2011ApJ...736..108P,2018ApJ...857...95M}. Thus, $t_{\rm fb}$ sets the characteristic transition between the rising and declining asymptotic regimes, but it is not the time at which the smooth fallback rate reaches its maximum.

{ We treat the fallback normalization $M_{\rm norm}$ as a free parameter and define
$\dot M_0=2M_{\rm norm}/(3t_{\rm fb})$. If the asymptotic $t^{-5/3}$
branch were extrapolated from $t_{\rm fb}$ to infinity, its integral would
equal $M_{\rm norm}$. For the complete smooth prescription in
Equation~(\ref{eq:fallback_rate}), the integral is instead approximately
$1.17456M_{\rm norm}$. This is supplied fallback mass, not necessarily
mass retained by the star. The maximum occurs at
$t_{\rm pk}=x_{\rm pk}t_{\rm fb}$, where
$x_{\rm pk}=(3/10)^{6/13}\simeq0.57368$, and
$\dot M_{\rm pk}\simeq0.58263\dot M_0$.
Both $M_{\rm norm}$ and $t_{\rm fb}$ are varied, and the normalization
and peak fallback rate are recalculated for each parameter vector.}

In the present model, we identify the mass inflow rate reaching the magnetospheric boundary directly with $\dot{M}_{\rm fb}(t)$. We do not introduce an additional viscous delay or smoothing timescale. Consequently, Equation~(\ref{eq:fallback_rate}) is used directly to calculate the magnetospheric radius, disk torque, and stellar spin evolution.
{ For the luminosity calculation, this is an effective prescription for
the inner inflow. Its identification with the ballistic mass-return
history without viscous filtering is an assumption, not a result of
a disk-transport calculation. The inferred $t_{\rm fb}$ and
$M_{\rm norm}$ retain this conditional interpretation.}

\subsection{Characteristic radii}

We denote the magnetar mass, radius, angular frequency, spin period, and equatorial surface dipole field by $M_\ast$, $R_\ast$, $\Omega_\ast=2\pi/P_\ast$, $P_\ast$, and $B_{\rm }$, respectively. The corresponding magnetic dipole moment is $\mu=B_{\rm }R_\ast^3$. The fallback disk is truncated at the magnetospheric radius

\begin{equation}
r_m(t)=\xi \left[\frac{9\mu^4}{4GM_\ast\dot{M}_{\rm fb}^2(t)}\right]^{1/7}
=\xi \left(\frac{3}{2}\right)^{2/7} G^{-1/7}M_\ast^{-1/7} \dot{M}_{\rm fb}^{-2/7}(t) B_{\rm }^{4/7}R_\ast^{12/7},
\label{eq:rm}
\end{equation}

where $\xi$ is an order-unity coefficient accounting for uncertainties in the geometry of the disk--magnetosphere interaction \citep[e.g., see][]{2005ApJ...634.1214L, 2008A&A...478..155B, 2013A&A...550A..99Z} and $G$ is the gravitational constant. Throughout this analysis, we fix the NS mass and radius to $M_\ast=1.4\,M_\odot$ and $R_\ast=12$ km, respectively, and adopt $\xi=1$. These quantities are held fixed in the Markov chain Monte Carlo (MCMC) analysis. Numerically, Equation~(\ref{eq:rm}) becomes $r_m\simeq
22.3\,{\rm km}\, \xi\, B_{{\rm }15}^{4/7} R_{\ast,12}^{12/7} M_{\ast,1.4}^{-1/7}
\dot{M}_{{\rm fb},-2}^{-2/7}$, where $B_{{\rm }15}=B_{\rm }/(10^{15}\ {\rm G})$, $R_{\ast,12}=R_\ast/(12\ {\rm km})$, $M_{\ast,1.4}=M_\ast/(1.4\,M_\odot)$, and $\dot{M}_{{\rm fb},-2}=\dot{M}_{\rm fb}/(10^{-2}\,M_\odot\,{\rm s}^{-1})$. We impose $r_m\geq R_\ast$ whenever the formal magnetospheric radius would fall below the stellar surface.

The corotation and light-cylinder radii are
\begin{equation}
r_{\rm co}=\left(\frac{GM_\ast}{\Omega_\ast^2}\right)^{1/3}
\simeq
16.8\,{\rm km}\,M_{\ast,1.4}^{1/3}P_{\ast,\rm ms}^{2/3},
\label{eq:rco}
\end{equation}
and
\begin{equation}
r_{\rm LC}= \frac{c}{\Omega_\ast} \simeq 47.7\,{\rm km}\,P_{\ast,\rm ms},
\label{eq:rlc}
\end{equation}
respectively, where $P_{\ast,\rm ms}=P_\ast/(1\ {\rm ms})$ and $c$ is the speed of light.

The fastness parameter measures the stellar angular velocity relative to the Keplerian angular velocity at the magnetospheric boundary
\begin{equation}
\omega_s= \frac{\Omega_\ast}{\Omega_K(r_m)} = \left(\frac{r_m}{r_{\rm co}}\right)^{3/2}.
\label{eq:fastness}
\end{equation}
The corresponding instantaneous corotation period at the inner edge,
defined by $\omega_s=1$, or equivalently $r_m=r_{\rm co}$, is
\begin{equation}
P_{\rm K}= \frac{2\pi}{\Omega_K(r_m)}= 2\pi
\left(\frac{r_m^3}{GM_\ast}\right)^{1/2}.
\label{eq:peq}
\end{equation}
Corotation cancels the disk torque only. Electromagnetic braking
remains nonzero, so this is not the equilibrium period of the total
torque; instantaneous spin equilibrium requires
$N_{\rm acc}=N_{\rm EM}$ in Equation~(\ref{eq:spin}).

Inside the light cylinder, $\omega_s<1$ corresponds to positive
disk torque, while $\omega_s>1$ corresponds to negative propeller torque
\citep{2011ApJ...736..108P}. The transition is at $\omega_s=1$.
For $r_m\geq r_{\rm LC}$, the adopted ejector prescription instead
sets the disk torque to zero. Penetration inside the light cylinder
can also enhance the open magnetic flux and electromagnetic output,
as described in Section~\ref{sec:spin_evolution}.

\subsection{Spin evolution and electromagnetic luminosity}
\label{sec:spin_evolution}

The stellar angular momentum evolves under the combined action of the disk and electromagnetic torques
\begin{equation}
I_\ast\frac{{\rm d}\Omega_\ast}{{\rm d}t}
=N_{\rm acc}-N_{\rm EM},
\label{eq:spin}
\end{equation}
 where $I_\ast=1.3\times10^{45}\ {\rm g\, cm^2}$ is the constant NS moment of inertia adopted throughout the spin-evolution calculation. This quantity is fixed rather than inferred from the X-ray luminosity fit \citep{2005ApJ...629..979L}. Here, $N_{\rm EM}>0$ denotes the magnitude of the electromagnetic braking torque, whereas the sign of $N_{\rm acc}$ is determined by the disk--magnetosphere interaction.  We write the disk torque as
\begin{equation}
N_{\rm acc}=
\begin{cases}
\dot{M}_{\rm fb}(GM_\ast r_m)^{1/2}n(\omega_s), & r_m<r_{\rm LC},\\
0, & r_m\geq r_{\rm LC}.
\end{cases}
\label{eq:nacc}
\end{equation}
and adopt the prescription $n(\omega_s)=1-\omega_s$ on the disk-coupled branch. The disk torque is switched off outside the light cylinder, defining the ejector branch.  This prescription produces a positive spin-up torque in the accretion regime, $\omega_s<1$, and a negative spin-down torque in the propeller regime, $\omega_s>1$ \citep{2011ApJ...736..108P}.  Thus, although denoted by $N_{\rm acc}$, the disk torque can be negative when the magnetosphere rotates faster than the inner disk. When the disk penetrates the light cylinder, it can open stellar magnetic flux that would otherwise remain closed, thereby enhancing the electromagnetic braking torque and luminosity \citep{2016ApJ...822...33P,2018ApJ...857...95M}. We describe this effect using the effective Y-point prescription $r_Y=\min\left[r_{\rm LC},\max(r_m,R_\ast)\right]$.  The electromagnetic braking torque is then
\begin{equation}
N_{\rm EM}
=\frac{L_{\rm EM}}{\Omega_\ast}
=\kappa_\chi\frac{\mu^2\Omega_\ast}{c\,r_Y^2},
\label{eq:nem}
\end{equation}
and the corresponding electromagnetic luminosity is
\begin{equation}
L_{\rm EM}
=\Omega_\ast N_{\rm EM}
=\kappa_\chi\frac{\mu^2\Omega_\ast^2}{c\,r_Y^2}
=L_{\rm dip}\left(\frac{r_{\rm LC}}{r_Y}\right)^2,
\label{eq:lem}
\end{equation}
where $L_{\rm dip} =\kappa_\chi\frac{\mu^2\Omega_\ast^4}{c^3}$ is the force-free spin-down luminosity of an isolated magnetar. We adopt $\kappa_\chi\simeq1+\sin^2\chi$, where $\chi$ is the angle between the magnetic and rotational axes \citep{Spitkovsky2006}, and use $\kappa_\chi=1$ for the fiducial aligned configuration. The magnetic obliquity is fixed in this calculation.

When $r_m\geq r_{\rm LC}$, the disk does not modify the open magnetic flux, $r_Y=r_{\rm LC}$, and Equation~(\ref{eq:lem}) reduces to the isolated force-free luminosity, $L_{\rm EM}=L_{\rm dip}$. For $R_\ast<r_m<r_{\rm LC}$, the disk compresses the closed magnetosphere and enhances the electromagnetic luminosity by $\frac{L_{\rm EM}}{L_{\rm dip}} =\left(\frac{r_{\rm LC}}{r_m}\right)^2$. If the formal magnetospheric radius falls below the stellar surface, the Y-point radius saturates at $r_Y=R_\ast$, yielding the maximum luminosity
\begin{equation}
L_{\rm EM,max}
=L_{\rm dip}\left(\frac{r_{\rm LC}}{R_\ast}\right)^2.
\end{equation}

An important distinction is that $L_{\rm EM}$ represents the electromagnetic power available to the relativistic outflow, whereas the net rate of change of the stellar rotational-energy reservoir also depends on the disk torque. For $E_{\rm rot}=\frac{1}{2}I_\ast\Omega_\ast^2$ and constant $I_\ast$, Equation~(\ref{eq:spin}) gives
\begin{equation}
-\frac{{\rm d}E_{\rm rot}}{{\rm d}t}
=\Omega_\ast(N_{\rm EM}-N_{\rm acc}).
\end{equation}
In the accretion regime, the positive disk torque replenishes part of the rotational energy extracted electromagnetically. In the propeller regime, the negative disk torque provides an additional channel for extracting rotational energy, which may be transferred to expelled material rather than to the electromagnetic wind. We therefore associate the observed early X-ray component with $L_{\rm EM}$ rather than with the net loss rate of the stellar rotational-energy reservoir.

\subsection{Model fitting}
\label{sec:model-fitting}

{ Our aim is to describe a smooth magnetar contribution beneath the variable
early X-ray emission. Individual excesses are treated statistically; their
dissipation mechanism is not fitted. No quasi-periodic oscillation (QPO) frequency or precession condition enters the likelihood or the priors.}

{ We fit the unabsorbed 0.3--10 keV flux directly, retaining 380 original XRT
bins with centres between 80 and 430 s after the BAT trigger. With
$t=(t_{\rm obs}-t_0)/(1+z)$, the physical prediction is
\begin{equation}
F_{\rm mag}(t_{\rm obs})=\frac{\epsilon_{\rm eff}L_{\rm EM}(t)}{4\pi D_L^2},
\qquad \epsilon_{\rm eff}=\frac{\eta_{\rm band}}{f_b}.
\label{eq:flux-current}
\end{equation}
At $z=1.411$, the corresponding source-frame band is 0.7233--24.11 keV;
$\eta_{\rm band}$ is defined in that matched band. No additional spectral
$K$-correction is applied to this unresolved component. For
$D_L=10.30$ Gpc, $4\pi D_L^2=1.26936\times10^{58}$ cm$^2$.
The spectral conversion embodied in the adopted unabsorbed XRT flux
product is retained. Photon indices are not additional fit parameters,
and their uncertainties are not separately propagated through a new
spectral fit. The inference is conditional on this conversion and the
supplied flux errors.
Each prediction is averaged over the original bin using five-point
Gauss--Legendre quadrature. Original flux errors are retained, with no
rebinning or selection of local minima. The later afterglow is neither
subtracted nor refitted, so additional emission near 400--430 s may include
an emerging external-shock contribution.

Let $r_i=\log_{10}F_i-\log_{10}F_{{\rm mag},i}$. We approximate the measured log-flux uncertainty by $\sigma_i=\tfrac12\left[\log_{10}(F_i+\Delta F_{i,+})- \log_{10}(F_i-\Delta F_{i,-})\right]$, where both error magnitudes are positive. The marginal residual density is
\begin{equation}
 f_i(r)=\int_0^\infty\frac{e^{-u/\tau}}{\tau}
 \frac{\exp[-(r-u)^2/(2\sigma_i^2)]}{\sqrt{2\pi}\sigma_i}\,du.
\end{equation}
The positive excess is defined in logarithmic flux. An excess $u$ corresponds
to additional linear flux $F_{{\rm mag},i}(10^u-1)$ at the marginal level.

For the correlated-residual inference, we transform these margins using
$z_i=\Phi^{-1}[H_i(r_i)]$, where $H_i$ is the marginal cumulative
distribution, and use the Gaussian copula $R_{ij}=a\exp[-|t_{{\rm obs},i}-t_{{\rm obs},j}|/\ell]+(1-a)\delta_{ij}$, and $\ln\mathcal L=\sum_i\ln f_i(r_i)-\tfrac12\ln|R|
 -\tfrac12\boldsymbol z^{\rm T}(R^{-1}-I)\boldsymbol z$.  The correlation timescale $\ell$ is in the observer frame. This working
model represents dependence in the observed residuals; it is not an exact
convolution of correlated physical emission and independent detector noise.

We sample five physical parameters, $P_0,B,t_{\rm fb},M_{\rm norm}$ and
$\epsilon_{\rm eff}$, and three residual parameters, $\tau,\ell,a$.
Priors are independent and uniform in the logarithmic coordinates listed
in Table~\ref{tab:fits}. The initial period is specified at the fixed
source-frame epoch $t_{\rm init}=10^{-3}$ s, with
$\log_{10}(P_0/{\rm ms})\in[0,3]$. The 1-ms lower cutoff excludes
sub-millisecond initial periods and is a working assumption, not a
universal equation-of-state-independent mass-shedding limit. The broad
1000-ms upper cutoff allows slowly rotating initial configurations and
subsequent fallback-driven spin-up; it is not a measured physical maximum.
No equation-of-state-dependent collapse or mass-shedding filter is imposed
during evolution.

For $\epsilon_{\rm eff}=\eta_{\rm band}/f_b$, we adopt
$\log_{10}\epsilon_{\rm eff}\in[-3,\log_{10}(81.2)]$; the positive
lower cutoff is a working bound for logarithmic sampling. For two uniform
emission cones of half-opening angle $\theta_j$, $f_b=1-\cos\theta_j$.
The fiducial assumptions $\eta_{\rm band}\leq1$ and
$\theta_j=9^\circ$ give
$\epsilon_{\rm eff}\leq[1-\cos(9^\circ)]^{-1}\simeq81.2$.
Values above unity therefore do not imply $\eta_{\rm band}>1$.
This is a conditional geometry assumption, not a universal efficiency
bound; the early X-ray beam need not share the afterglow opening angle.
The separate descriptive candidate reaches the adopted ceiling, and its
interpretation remains conditional on this bound.

Four independent replica-exchange runs each contain ten inverse temperatures
spanning $\beta=1$ to $10^{-3}$. We exclude the 1000 warm-up steps and
retain 60000 cold-chain production states per run, for 240000 correlated
posterior samples. Local proposals and frozen global independence proposals
use the full Hastings correction; the global proposal combines six Student
$t$ components and a uniform-box component. Production proposals are not
adapted. Rank-normalized split $\widehat R$ and bulk/tail effective sample
sizes are computed across the four independent cold runs. The retained cold-chain samples, proposal settings and numerical
validation results are available from the corresponding author upon
reasonable request, as specified in the Data Availability statement.}

{ \subsubsection{A separate descriptive estimate and temporal-block uncertainty.}
To construct a candidate curve closer to the smooth support of the data,
we also minimize the marginal objective $-\sum_i\ln f_i(r_i)$, equivalently
setting $R=I$ for this estimator. This changes the statistical treatment,
not the physical flux calculation. The six parameters
$P_0,B,t_{\rm fb},M_{\rm norm},\tau,\epsilon_{\rm eff}$ are optimized
within the same corresponding bounds. Their uncertainty is explored by
nonoverlapping temporal-block pairs bootstrap, rather than by interpreting
the independent-bin objective as a calibrated joint likelihood.

Blocks of 10, 20 and 40 s are anchored at observer time 80 s and assigned
by bin centre. Each realization resamples the original number of blocks
with replacement, retaining the times, fluxes and errors of every selected
bin and assigning its block multiplicity as a weight. We retain 100 refits
per block duration, including boundary solutions. Each refit uses four
dispersed starts and a local refinement. Percentiles of these refits provide
exploratory uncertainty ranges; they are not posterior credible intervals.
Resampling preserves dependence within a block, but not across blocks.
The 40-s case has only nine original blocks, so block-length sensitivity
is reported explicitly.  The adopted block lengths explore sensitivity to short-timescale dependence; they do not establish robustness to correlations extending beyond 40 s.

\subsection{Dynamical quantities and statistical interpretation}
\label{sec:dynamical-scope}
For each joint physical parameter vector, we evolve the spin and evaluate
$r_{\rm m}$, $r_{\rm co}$ and $r_{\rm LC}$ using the same trajectory.
We then form radius ratios and indicator functions for the relevant
inequalities. This preserves parameter dependence and avoids interpreting
a collection of marginal medians as a single solution. Pointwise
credible intervals describe uncertainty at each epoch; they are not
simultaneous credible bands for an entire trajectory.

The probabilities of a radius inequality at a specified epoch and at
every epoch of a finite evaluation grid are different summaries.
The grid is the union of 241 uniformly spaced observer-frame epochs
from 80 to 430 s (spacing $350/240$ s) and the exact epochs 200 and
400 s, giving 243 evaluation times. Neither
is inferred by inspecting the overlap of marginal radius intervals.
The propagated probabilities inherit the assumptions of the fitted
residual model; their interpretation is subject to the predictive
limitations in Section~\ref{sec:correlated-posterior}.
No radius defined by matching a timing period is included in this
propagation. The numerical results are given in
Section~\ref{sec:joint-dynamics}.

\section{Results}
\label{sec:results}

\subsection{X-ray flux descriptions and parameter constraints}
\label{sec:xray-results}

We fit the unabsorbed observer-frame 0.3--10 keV XRT flux over
$80\leq t_{\rm obs}-t_0\leq430$ s with the physical model and residual
prescriptions of Section~\ref{sec:model-fitting}. The aim is to describe
a possible smooth magnetar contribution beneath the variable emission.
No QPO frequency or Lense--Thirring condition enters the fit.

Table~\ref{tab:fits} and Figure~\ref{fig1:mcmc} summarize the
correlated-residual posterior. Its medians are $P_0=16.8$ ms,
$B=1.56\times10^{15}$ G, $t_{\rm fb}=54.8$ s in the source frame,
and $\epsilon_{\rm eff}=2.34$. In particular, the central 68 per cent
range of $P_0$ is 2.76--286 ms, so this analysis does not determine a
precise initial spin period. The physical flux bands in
Figure~\ref{fig:flux-current}(a) are obtained from joint posterior draws,
not from a trajectory evaluated at the marginal median parameters.

The separate marginal descriptive candidate in Table~\ref{tab:bootstrap}
has $P_0=4.25$ ms, $B=3.04\times10^{15}$ G,
$t_{\rm fb}=143.1$ s in the source frame, $M_{\rm norm}=0.0090M_\odot$
and $\epsilon_{\rm eff}=81.2$. It is shown in
Figure~\ref{fig:flux-current}(b) and as the solid magenta curve in
Figure~\ref{fig2:radii}. Its temporal-block ranges are distinct from the
posterior credible intervals. Both descriptions use the same conversion
$F_{\rm mag}=\epsilon_{\rm eff}L_{\rm EM}/(4\pi D_L^2)$.

For the descriptive candidate, the fallback-rate maximum occurs at
$(1+z)x_{\rm pk}t_{\rm fb}\simeq198$ s after the trigger. The smooth
magnetar flux instead has its largest value within the fitted interval
at 80 s and subsequently declines. The total observed peak near 180 s
is not attributed to a maximum of this smooth component; the distinction
and the explicit flux normalization are detailed in
Section~\ref{sec:flux-normalization}.

Figure~\ref{fig2:radii} also shows the broader XRT evolution and the
BAT count rate for temporal context. Neither the BAT data nor the XRT
measurements outside 80--430~s enter the magnetar fit. Possible overlap with external-shock
emission near 400--430 s remains part of the interpretation of the
selected interval.}

\subsection{Spin evolution and disk--magnetosphere regime}
\label{sec:spin-results}

{ Figure~\ref{fig:dynamics-current} shows the stellar spin period and
characteristic-radius ratios for the two flux descriptions. Its left
column propagates joint posterior draws, while its right column shows
the descriptive candidate with the ranges from the 20-s temporal-block
refits. These summaries must not be combined into a single representative
trajectory.

At $t_{\rm obs}-t_0=200$ s, the descriptive candidate has
$P_*=4.816$ ms, $r_{\rm m}=234.8$ km and $r_{\rm LC}=229.8$ km.
The disk is still outside the light cylinder, so $N_{\rm acc}=0$ and
$r_Y=r_{\rm LC}$ under the adopted prescription. It enters the light
cylinder near observer time 255 s. The subsequent disk-coupled phase
has $r_{\rm m}>r_{\rm co}$ and a negative propeller torque, producing
additional spin-down and a steeper decline in the magnetar flux.

The correlated posterior permits a broader set of dynamical histories.
At 200 s, the joint-draw fractions satisfying $r_{\rm m}<r_{\rm co}$
and $r_{\rm m}<r_{\rm LC}$ are approximately 0.55 and 0.83,
respectively. These are separate conditions, not a joint probability
of sustained accretion. The full propagated ranges and finite-grid
conditions are reported in Section~\ref{sec:joint-dynamics}.
The revised analysis therefore does not imply that positive disk torque
or a fixed luminosity enhancement persists throughout the fitted interval.

\subsection{Limits of geometrical inference}
\label{sec:geometry-limits}
The joint propagation in Section~\ref{sec:joint-dynamics} permits
several disk--magnetosphere histories. A radius ordering obtained from
one illustrative trajectory does not establish its posterior probability
or its persistence throughout the XRT interval. In particular, the
inference does not establish continuous positive disk torque.

No timing-based Lense--Thirring radius or global precession period is
inferred in this analysis. The local timescale estimates in
Section~\ref{sec:discussion_lt} do not identify a precessing region.
We do not propagate an instrument-dependent timing-centroid
range, assign a probability to $r_{\rm m}<r_{\rm LT}$, or identify a
coherently precessing annulus. The previously considered timing-based
radius hierarchy is not retained as a physical constraint or a result.

\subsection{Correlated-residual posterior}
\label{sec:correlated-posterior}
Table~\ref{tab:fits} and Figure~\ref{fig1:mcmc} describe the same retained
MCMC calculation, with a free fallback normalization and the hybrid torque
prescription. The posterior medians are $B=1.56\times10^{15}$ G,
$t_{\rm fb}=54.8$ s in the source frame and $\epsilon_{\rm eff}=2.34$.
Their central 68 per cent ranges are respectively
$(0.50$--$4.01)\times10^{15}$ G, 27.9--284.7 s and 0.281--21.0.
The initial period has median 16.8 ms and central 68 per cent range
2.76--286 ms. Approximately 29 per cent of the samples have $P_0>100$ ms.
The restricted plotting range in Figure~\ref{fig1:mcmc} changes neither
these intervals nor the probability assigned to its omitted tail.
A full-range marginal distribution is provided in
Appendix~\ref{sec:full-p0}, Figure~\ref{fig:p0-full}, using the same
240000 retained cold-chain samples.

The retained runs satisfy the adopted numerical sampling criteria: the
largest rank-normalized split $\widehat R$ is 1.0072; the smallest
continuous-parameter bulk and tail effective sample sizes are
approximately 732 and 444, respectively. For $P_0$, $\widehat R=1.0028$,
with bulk and tail effective sample sizes approximately 1474 and 2755.
Increasing the numerical precision changes the tested log-likelihoods
by at most $4.70\times10^{-4}$. These checks address numerical sampling.
The 240000 retained states are correlated, and the heated replicas are
not treated as independent posterior chains. Numerical convergence
does not establish model adequacy or a unique physical decomposition.
The component-flux bands in Figure~\ref{fig:flux-current}(a) are not
posterior predictive bands for the total observed flux.

The residual correlation timescale has median 286 s and central
68 per cent range 247--308 s. About 96 per cent of the samples lie in
the upper 5 per cent of its logarithmic prior interval, whose upper bound
is $10^{2.5}$ s. The inferred timescale consequently remains sensitive
to that bound. The dependence amplitude has median $a=0.9973$.
The likelihood can assign a sustained positive excess to the variable
contribution, leaving the magnetar component faint. For the propagated
draws in Figure~\ref{fig:flux-current}(a), the median across bins of the
ratio of the posterior median magnetar flux to the measured flux is
0.029. This ratio is a descriptive comparison, not a fluence fraction
or a goodness-of-fit probability.

We assess the statistical description of the total observed flux through
1000 posterior predictive replications, using the same joint posterior
draws propagated for the dynamics, with 250 draws from each independent
cold run. For each draw, we combine the bin-averaged physical flux with
a new residual realization from the fitted Gaussian copula and
exGaussian margins at the original 380 bin centres between 80 and
430 observer-frame seconds. The measured log-flux error estimates are
held fixed and included once through the marginal distributions.
The replicated residual path is not conditioned on the observed path.
This calculation requires no further fitting or MCMC. Appendix~\ref{app:ppc}
gives the replication prescription, the numerical summaries in
Table~\ref{tab:ppc} and the diagnostic Figure~\ref{fig:ppc}.

Three summary statistics were specified before generating the
replications: the sample standard deviation of logarithmic flux, its
maximum, and the root-mean-square logarithmic flux difference for pairs
separated by 10--40 observer-frame seconds. The last statistic uses
11907 pairs with equal pair weights. For each statistic, $p_B$ denotes
the fraction of replicated values greater than or equal to the observed
value. These posterior predictive tail fractions are not
frequentist-calibrated $p$-values and are not interpreted using a formal
rejection threshold.

The observed log-flux standard deviation is 0.593 dex, compared with a
replicated median of 0.277 dex; 48 of 1000 replications reach or exceed
the observed value ($p_B=0.048$). The observed peak flux is
$2.49\times10^{-7}\ \mathrm{erg\,cm^{-2}\,s^{-1}}$, reached or exceeded
by 32 replications ($p_B=0.032$). The observed root-mean-square
logarithmic flux difference on 10--40 s separations is 0.310 dex,
compared with a replicated median of 0.149 dex and $p_B=0.071$.
As an additional descriptive check, only 79 of 380 measured fluxes
(20.8 per cent) lie within the central 90 per cent pointwise predictive
intervals, while the remaining 301 lie above their upper limits.
This inclusion fraction is not an independent-bin coverage test: the
bins are dependent, the model was fitted to these same data, and the
intervals are pointwise rather than simultaneous.

These checks reveal shortcomings in the current correlated-residual
prescription, whose replicated light curves typically have lower flux
levels, smaller overall dispersion and lower peaks than observed.
The reported posterior summaries therefore remain conditional on a
statistical prescription that does not adequately reproduce these
features of the data. They should not be interpreted as validated
uncertainty estimates for an independently identified magnetar
component. This assessment concerns the combined physical-flux and
statistical-residual description; it does not by itself exclude a
magnetar contribution. It also does not validate the separate
descriptive candidate or its temporal-block bootstrap ranges.}

{ \subsection{A separate descriptive curve and block uncertainty}
Figure~\ref{fig:flux-current}(b) shows the candidate from the marginal
objective and its temporal-block uncertainty, as defined in
Section~\ref{sec:model-fitting}. The original-data candidate has
$P_0=4.25$ ms, $B=3.04\times10^{15}$ G, $t_{\rm fb}=143$ s,
$M_{\rm norm}=8.99\times10^{-3}M_\odot$, $\tau=0.342$ dex and
$\epsilon_{\rm eff}=81.2$. The latter reaches the adopted ceiling.
The curve describes a smooth contribution below the variable data;
the median binwise model-to-data flux ratio is 0.542. Its uncertainty
comes from the bootstrap refits, not from the MCMC in Figure~\ref{fig1:mcmc}.

Table~\ref{tab:bootstrap} displays the sensitivity to 10-, 20- and 40-s
blocks. The empirical 68 per cent ranges of $P_0$ are 3.88--4.64,
3.28--4.37 and 1.00--4.66 ms, respectively. In the 40-s case, 18 of
100 refits reach approximately the 1-ms lower search boundary, and three
reach the long-period boundary near 1000 ms. These three are retained
although they lie outside the central percentile range. They occur in
resamples lacking the late observations after approximately 280--320 s.
The 10-, 20- and 40-s cases place 86, 77 and 65 per cent of refits
within one per cent of the efficiency ceiling, respectively.
The reported ranges therefore characterize a conditional descriptive
estimate, not an independently established narrow birth-spin interval.

\subsubsection{Sensitivity to the upper time limit.}
We repeat the marginal descriptive optimization using the 361 original
bins with centres in 80--400 s, compared with 380 in 80--430 s.
The model, six fitted parameters, bounds, original bin widths and
errors, and five-point bin averaging are unchanged. Both windows are
optimized from five starting points followed by a local refinement.
The 80--430 s refit reproduces the retained candidate. The comparison
is given in Table~\ref{tab:endpoint}; it uses the actual data with unit
weights, rather than block resamples. No additional MCMC or endpoint-specific
bootstrap intervals are computed.

The shorter-window fit changes $P_0$, $B$, $t_{\rm fb}$ and
$M_{\rm norm}$ by approximately $-3.4$, $-8.4$, $-14.2$ and $-10.3$
per cent, while $\epsilon_{\rm eff}$ remains at its upper bound.
Its flux is $1.919\times10^{-8}$ at 200 s and
$2.027\times10^{-9}$ erg cm$^{-2}$ s$^{-1}$ at 400 s, compared with
$1.902\times10^{-8}$ and $1.810\times10^{-9}$ for the reference
candidate. The maximum relative curve difference on a 1-s grid over
80--400 s is 12.0 per cent, using the 80--430 s curve as the denominator. This is a point-estimate sensitivity check,
not evidence that the last 19 bins are free of afterglow emission.

As a diagnostic for the correlated analysis, we reweight the 1000
already propagated joint draws by
$w_j\propto\mathcal L_{80-400,j}/\mathcal L_{80-430,j}$ with unchanged
priors and the appropriate retained-bin copula. The normalized weights
give $1/\sum_jw_j^2\simeq515$, with maximum weight 0.0216. This
quantifies weight concentration, not a chain-autocorrelation effective
sample size. The initial-period distribution remains broad, while the
84th percentile of $t_{\rm fb}$ changes from approximately 292 s in
the unweighted 1000-draw subset to approximately 910 s after reweighting.
Thus the tail of the correlated inference is sensitive to the endpoint.
These subset-based diagnostics do not constitute a separately converged
posterior or replace the full-chain reference results.

\subsection{Flux normalization and the observed peak}
\label{sec:flux-normalization}
The descriptive curve in Figure~\ref{fig:flux-current}(b) and the
solid magenta curve in Figure~\ref{fig2:radii} use the same candidate
parameters from Table~\ref{tab:bootstrap}. At observer time 200 s
($t=82.95$ s in the source frame), this candidate has $P_*=4.816$ ms, $r_{\rm m}=234.8$ km and
$r_{\rm LC}=229.8$ km. It is still on the ejector branch, so
$r_Y=r_{\rm LC}$, and its electromagnetic luminosity is
$L_{\rm EM}=2.973\times10^{48}$ erg s$^{-1}$. The plotted flux is
\begin{equation}
F_{\rm mag}(200\,\mathrm{s})=
\frac{81.2\,(2.973\times10^{48})}{1.269\times10^{58}}
=1.902\times10^{-8}\;\mathrm{erg\,cm^{-2}\,s^{-1}}.
\end{equation}
There is no additional plotting normalization or repeated beaming factor.
For comparison, the correlated MCMC gives a pointwise median flux of
$7.23\times10^{-10}$ erg cm$^{-2}$ s$^{-1}$ at this epoch, with central
68 per cent interval $(3.72$--$12.04)\times10^{-10}$.
This median is calculated from joint-draw fluxes, not from the product
of marginal medians of $\epsilon_{\rm eff}$ and $L_{\rm EM}$.

The candidate's fallback-rate maximum occurs near 198 s in the observer
frame. Its electromagnetic flux is largest at $t_{\rm obs}-t_0=80$ s,
the beginning of the fitted 80--430 s interval, and subsequently declines.
This in-interval maximum precedes the observed total-flux peak near 180 s;
we do not infer the epoch of a global maximum outside the fitted interval.
The maximum of the mass-supply rate need not coincide with that of the
electromagnetic luminosity, which also depends on the stellar spin and
Y-point radius. For constant $\epsilon_{\rm eff}$, simultaneous
spin-down and expansion of $r_{\rm m}$ on the disk-coupled branch
produce a declining luminosity, not a delayed smooth-component peak.

The peak near 180 s includes additional variable emission, represented
statistically by the excess component. Changes in outflow dissipation
are one possible source of variability, but no energy-conserving
joint model of the smooth component and the excesses is constructed.
In particular, the descriptive candidate already reaches approximately
unit band efficiency for the adopted beaming geometry. Additional
positive flux cannot then be supplied merely by increasing the
conversion of the same instantaneous $L_{\rm EM}$ with unchanged
beaming. The excesses are not necessarily powered by that same
instantaneous reservoir; other energy sources, prior storage and
different geometries are possibilities, not conclusions of this fit.
No additional viscous delay or time-dependent X-ray efficiency is
introduced to align the smooth model with the observed peak.
The candidate enters the light cylinder near observer time 255 s;
the subsequent propeller torque produces a steeper decline.

\subsection{Joint propagation of the dynamics}
\label{sec:joint-dynamics}
We propagate 1000 joint posterior draws, selected uniformly without
replacement with 250 from each independent cold run, using a fixed seed.
Figure~\ref{fig:dynamics-current} separates the resulting credible bands
from the empirical bands obtained by propagating all 100 refits of the
20-s bootstrap case. In each case the same joint parameter vector is
used for the spin and all radii. Ratios are evaluated per trajectory,
rather than formed from independently summarized marginal radii.

At observer time 200 s, the correlated posterior gives $P_*=6.1$ ms
with central 68 per cent interval 2.3--17.1 ms, $r_{\rm m}=85$ km
(32--203 km), $r_{\rm co}=56$ km (29--111 km), and $r_{\rm LC}=290$ km
(109--817 km). These are separate marginal summaries, not one physical
trajectory. The following probabilities are conditional on the
correlated-residual prescription, whose predictive shortcomings are
reported in Section~\ref{sec:correlated-posterior}. They are estimated
from the fractions of joint draws satisfying each inequality:
\begin{equation}
\begin{array}{c|cc}
t_{\rm obs}-t_0 & \Pr(r_{\rm m}<r_{\rm co}) & \Pr(r_{\rm m}<r_{\rm LC})\\ \hline
200\,\mathrm{s} & 0.552 & 0.834\\
400\,\mathrm{s} & 0.287 & 0.915
\end{array}
\end{equation}
The fractions satisfying each condition at every evaluated epoch in
80--430 s are 0.277 and 0.788, respectively. These are separate
probabilities, not the probability of their intersection. They refer
to the evaluated finite grid and do not exclude crossings between
grid points. Positive disk torque requires both inequalities in the
adopted prescription, so these results do not establish persistent
accretion spin-up throughout the interval. The bands in
Figure~\ref{fig:dynamics-current} are pointwise; they are not
simultaneous credible bands for entire trajectories.

The descriptive candidate instead has $r_{\rm m}/r_{\rm co}=4.91$
and $r_{\rm m}/r_{\rm LC}=1.02$ at 200 s, changing to 2.57 and 0.37
at 400 s. Thus its late disk-coupled phase lies in the propeller regime.
About one per cent of the 1000 propagated posterior trajectories reach
$P_*<1$ ms somewhere on the evaluated grid; no such trajectory is
removed from the reported distributions. The fixed stellar-structure
model therefore supplies conditional dynamical estimates rather than
an equation-of-state-dependent survival or rotation test.}

\section{Discussion}
\label{sec:discusion}

 \subsection{A fallback-accreting magnetar as the early X-ray engine}
\label{sec:discussion-engine}

The bright, variable early X-ray episode motivates testing a smooth
contribution from continued central-engine activity. Fallback can modify
the spin and electromagnetic output of a magnetar; this framework has
also been applied to GRB~211211A and GRB~250704B/EP250704a
\citep{2026MNRAS.549ag949F,2026MNRAS.550g1295F}. These events illustrate
the mechanism across different GRB classes, rather than providing direct
population analogues of GRB~241030A.

{ The normalization inferred here depends on the treatment of the variable
excesses. The correlated-residual posterior permits a faint smooth
component, whereas the separate descriptive candidate follows a higher
support beneath the data. At 200 s their respective fluxes are
$7.23\times10^{-10}$ and $1.902\times10^{-8}$ erg cm$^{-2}$ s$^{-1}$.
The same physical luminosity-to-flux conversion is used in both cases;
the difference is not a plotting normalization.

The descriptive candidate's fallback rate peaks near 198 s in the
observer frame, but its smooth flux is already declining from the
beginning of the fitted interval at 80 s. The observed peak near 180 s
therefore includes additional variable emission. It is not reproduced
as a delayed smooth magnetar maximum, and no viscous delay or variable
band efficiency is invoked. The candidate remains on the ejector branch
at 200 s and subsequently enters a disk-coupled propeller phase near
255 s. These properties supersede the earlier interpretation of sustained
accretion spin-up through the fitted interval.

The effective normalization $\epsilon_{\rm eff}=\eta_{\rm band}/f_b$
combines the matched-band efficiency and collimation. The descriptive
candidate reaches the fiducial ceiling 81.2, corresponding approximately
to unit band efficiency for two uniform cones of half-opening angle
$9^\circ$. The angle is motivated by the afterglow context
\citep{Ducoin2026}, but identifying it with the early X-ray beam remains
an assumption. Neither analysis independently determines the efficiency
and beaming fraction, and the resulting physical interpretation remains
conditional on this geometry.
For the descriptive candidate, approximately unit band efficiency
leaves no comparable unused instantaneous electromagnetic power
under the same geometry. The energy budget of the smooth component
and the positive excesses has not been closed jointly; the statistical
excess prescription does not establish their energy source.

The later afterglow is not subtracted or refitted. Emission near
400--430 s may include both central-engine and external-shock
contributions, so the observed light curve does not establish a unique
physical decomposition.}

\subsection{Lense--Thirring precession and alternative modulation mechanisms}
\label{sec:discussion_lt}
\label{sec:variability-discussion}

The smooth X-ray component and the superposed variability probe
different aspects of the central engine. Our calculation examines
whether fallback-modified magnetar activity can sustain an underlying
luminosity, whereas the individual excesses may reflect changes in
dissipation, disk--magnetosphere coupling, or outflow geometry.
The inferred spin evolution and characteristic radii provide a
physical context for discussing these mechanisms. No modulation
period is used as an observational constraint in the present analysis,
and the residual correlation timescale $\ell$ is not interpreted
as a physical oscillation period.

Lense--Thirring precession of a tilted inner flow offers a possible
geometrical mechanism for variability
\citep{Fragile2007,Ingram2009}. If the flow can communicate the
differential precessional torque and maintain a coherent orientation,
its motion could change the direction of an associated outflow.
Relativistic beaming could then translate angular variations into
changes in the observed flux. The resulting behaviour depends on
the radial extent and surface-density profile of the flow, its
tilt, and its coupling to the emitting region. These quantities
are not determined by our luminosity model.

{  The updated dynamics do not establish a unique precessing geometry.
At observer time 200~s, corresponding to a source-frame epoch of
approximately 83~s, the descriptive candidate has
$P_*\simeq4.82$~ms, $r_{\rm m}\simeq235$~km,
$r_{\rm co}\simeq48$~km, and $r_{\rm LC}\simeq230$~km.
The ordering $r_{\rm co}<r_{\rm LC}<r_{\rm m}$ places this
realization on the ejector branch of the adopted prescription.
It enters a disk-coupled propeller phase near observer time
255~s. The correlated posterior permits a broader range of
dynamical histories, as shown in
Figure~\ref{fig:dynamics-current}; its marginal radius summaries
do not define a single trajectory. Consequently, neither a
precession radius nor a coherently precessing outer boundary
can be inferred from these results alone.

Nevertheless, characteristic timescales can be estimated without
identifying an observed periodicity. We use the descriptive trajectory
for the following illustrative calculations; these are not posterior
constraints. Keeping only the weak-field Lense--Thirring term gives
\begin{equation}
\Omega_{\rm LT}(r)=\frac{2GI_*\Omega_*}{c^2r^3},\qquad
P_{\rm LT,obs}(r)=(1+z)\frac{c^2r^3P_*}{2GI_*}.
\label{eq:lt-local-scale}
\end{equation}
For $I_*=1.3\times10^{45}$ g cm$^2$, evaluation at the formal
magnetospheric radius gives $P_{\rm LT,obs}(r_{\rm m})\simeq
7.8\times10^2$ s at observer time 200~s. At 400~s, the same
trajectory has $P_*\simeq14.2$ ms and $r_{\rm m}\simeq252$ km,
giving $P_{\rm LT,obs}(r_{\rm m})\simeq2.9\times10^3$ s.
For a coherently tilted, prograde disk extending outwards from
$r_{\rm in}=r_{\rm m}$, the rigid-precession frequency is an
angular-momentum-weighted average of the local frequencies
\citep{Ingram2009,Motta2018Timescales}. With Lense--Thirring as the
only precessional torque, its period cannot be shorter than the
local period at the inner edge. This descriptive realization
therefore does not favour Lense--Thirring precession as a clock
with a period of tens of seconds. This statement is conditional
on the assumed inner boundary and does not exclude other
trajectories. Magnetic precessional torques and stellar-quadrupole
contributions are not included in this estimate.

An illustrative communication test can also be made for a hypothetical,
nearly Keplerian annulus with $r_{\rm in}=r_{\rm m}$ and
$r_{\rm out}=2r_{\rm m}$. In the small-warp, bending-wave limit,
$\alpha<h\equiv H/r$, the propagation speed is approximately
$c_{\rm s}/2$ \citep{Lubow2002Warp}. Taking constant $h$ and
$c_{\rm s}=h\sqrt{GM_*/r}$ gives the source-frame crossing time
\begin{equation}
t_{\rm wave}\simeq\int_{r_{\rm in}}^{r_{\rm out}}
\frac{2\,\mathrm{d}r}{c_{\rm s}}
=\frac{4\bigl(r_{\rm out}^{3/2}-r_{\rm in}^{3/2}\bigr)}
{3h\sqrt{GM_*}}.
\label{eq:warp-illustrative}
\end{equation}
At observer time 200~s, $\alpha=0.01$ and $h=0.1$ give
$t_{\rm wave}\simeq0.20$ s; varying $h$ from 0.05 to 0.2 gives
approximately 0.41--0.10 s. These source-frame times are short
compared with both the local source-frame Lense--Thirring period
of approximately 323 s and the differential-twist timescale
$|\Omega_{\rm LT}(r_{\rm in})-\Omega_{\rm LT}(r_{\rm out})|^{-1}
\simeq59$ s, defined for a one-radian differential twist.
Thus, communication is sufficiently fast in this idealized
annulus. This necessary timescale condition does not establish
a coherent precessing disk: its assumed boundaries and thickness
are not inferred, the reference realization is on the ejector
branch, and neither warp damping nor magnetic stresses are evolved.
No comparison with an assumed 10.4-s modulation period is used.}

\subsubsection{Free precession of the magnetar}

Free precession of a deformed magnetar provides another possible
geometrical mechanism \citep{Shan2025}.
For a biaxial star, its approximate source-frame period is
\begin{equation}
P_{\rm fp}\simeq
\frac{P_*}{|\epsilon_{\rm fp}\cos\theta_{\rm w}|},
\end{equation}
where $\epsilon_{\rm fp}$ is the stellar ellipticity and
$\theta_{\rm w}$ is the wobble angle. The luminosity calculation
constrains the spin only conditionally on the adopted physical
and residual models; it does not determine the stellar deformation,
wobble angle, or damping time. For illustration, at observer time
200~s the descriptive spin gives
$P_{\rm fp,obs}\simeq116\,[|\epsilon_{\rm fp}\cos\theta_{\rm w}|/
10^{-4}]^{-1}$ s. The normalization $10^{-4}$ is an assumed
deformation--orientation product, not a measurement or an additional
prior. Without an independently established modulation period or
a deformation model, this scaling does not establish free precession
as the origin of the variability.

\subsubsection{Cyclic disk--magnetosphere interaction}

Mass accumulation and release near the magnetospheric boundary
can also produce variability, as illustrated by trapped-disk and
magnetospheric-gating calculations
\citep{2010MNRAS.406.1208D,2012MNRAS.420..416D}.
A characteristic local viscous timescale is
\begin{equation}
t_{\rm visc,obs}\sim(1+z)\alpha^{-1}
\left(\frac{H}{r}\right)^{-2}
\left(\frac{r_{\rm m}^{3}}{GM_*}\right)^{1/2},
\end{equation}
where $\alpha$ is the viscosity parameter and $H/r$ is the
relative disk thickness. An actual accumulation--release cycle
need not equal this local timescale.

{  At observer time 200~s, the descriptive value $r_{\rm m}\simeq235$ km
gives $t_{\rm visc,obs}\simeq201\,(\alpha/0.01)^{-1}
[(H/r)/0.1]^{-2}$ s. Thus, for $H/r=0.1$ and
$\alpha=0.01$--$0.1$, the illustrative local transport time spans
approximately 20--200 s. This range reflects selected disk parameters,
not an uncertainty interval. Because the reference realization is
on the ejector branch, it is a formal transport estimate at
$r_{\rm m}$, not evidence for active magnetospheric gating there.
For $\alpha=0.01$ and $H/r=0.1$, the corresponding source-frame
time is approximately 83 s, comparable to the source-frame age at
this epoch. This example therefore does not justify instantaneous
transport of returning material to the magnetosphere. The
circularization radius and radial transport are not calculated;
the effective inner-inflow prescription in Section~3.1 is not
validated by this local estimate.

The evolution of $r_{\rm m}/r_{\rm co}$ and
$r_{\rm m}/r_{\rm LC}$ in Figure~\ref{fig:dynamics-current}
identifies changes in the interaction regime within our
prescription. In particular, the descriptive candidate develops
a disk-coupled propeller phase, while the correlated posterior
allows more diverse histories. However, the present model does
not evolve mass storage at the inner boundary or determine
$\alpha$, $H/r$, and the interaction-region width. It therefore
does not predict a cycle period or establish magnetospheric
gating as the origin of the observed excesses.}

\subsubsection{Implications for the early X-ray variability}

Magnetic reconnection, shocks, and changes in radiative efficiency
or outflow orientation remain possible sources of the variable
emission. Likewise, a gamma-ray quiescent interval does not by
itself measure the fallback supply: our model assumes a continuous
mass-return history, without identifying the gamma-ray count rate
with the instantaneous inflow. For the descriptive candidate,
that supply peaks near observer time 198~s. This assumed continuity
does not establish uninterrupted accretion onto the star or
coherent precession.

{ These estimates give distinct characteristic timescales under explicit
assumptions. In the descriptive realization, the Lense--Thirring
timescale at the assumed inner edge is much longer than tens of seconds,
whereas illustrative viscous and free-precession scales can be shorter.
This comparison does not identify a variability mechanism or establish
that the assumed disk or stellar deformation is realized. A predictive
precession or cyclic-accretion interpretation would require independent timing
constraints and a model for the evolution and coherence of the
relevant emitting structure. Our conclusions about the smooth
fallback-magnetar contribution do not rely on a QPO detection.}

\subsection{Model assumptions and scope}
\label{sec:limitations}
We fix $M_\ast$, $R_\ast$, $I_\ast$, $\xi$ and $\kappa_\chi$ to the
stated fiducial values, while sampling $P_0$, $B$, $t_{\rm fb}$,
$M_{\rm norm}$ and $\epsilon_{\rm eff}$ together with three residual
parameters. The fallback normalization is free. The effective efficiency
combines band conversion and beaming, and its adopted ceiling is
conditional on a fiducial emission geometry.

{ The signed disk torque operates only inside the light cylinder; outside
it the disk torque is set to zero. Together with the Y-point prescription,
this defines a simplified evolution rather than a hydrodynamical model
of the disk, outflow, or magnetospheric interface. We do not fit individual
excesses or infer the mechanism that produces them.

The copula model is a working description of the residual distribution
and temporal dependence. Its prior-sensitive correlation timescale and
the broad initial-period distribution limit physical interpretation.
The posterior predictive assessment in Section~\ref{sec:correlated-posterior} reveals additional
shortcomings: replicated light curves typically have lower flux levels,
smaller overall dispersion and lower peaks than observed. Numerical
convergence therefore does not establish an adequate statistical
description of the total observed emission. The posterior summaries
remain conditional on this residual prescription and should not be
interpreted as validated uncertainty estimates for an independently
identified magnetar component. This assessment does not by itself
exclude a magnetar contribution, nor does it validate the separate
descriptive candidate or its temporal-block bootstrap ranges.
The supplied flux conversion is held fixed, and no separate spectral
uncertainty propagation is included.

The model is applied to the early XRT interval. The possible overlap with
afterglow emission near 400--430 s remains a separate uncertainty. The
endpoint comparison in Table~\ref{tab:endpoint} gives moderate changes
in the descriptive point estimate, while the reweighting diagnostic
shows sensitivity in the tail of the correlated inference. These
assumptions allow a conditional descriptive interpretation but do not
establish a unique emission decomposition, a persistent accretion regime,
or a precession origin of the variability.
\subsection{A descriptive magnetar contribution}
The fallback-modified spin evolution permits a smooth contribution beneath
the early XRT variability. Its inferred normalization depends on how the
additional emission and temporal dependence are represented. The
correlated-residual posterior and the marginal descriptive estimate
therefore answer different statistical questions. The former quantifies
conditional parameter uncertainty under an explicit copula; the latter
constructs a candidate smooth support and examines its stability under
temporal-block resampling. The posterior intervals and the bootstrap ranges are reported separately
for their respective curves.

The descriptive candidate uses a millisecond initial period and a field of
order $10^{15}$ G. This establishes a possible model realization, rather
than a unique identification of the central engine or a precise birth-spin
measurement. In particular, its effective efficiency reaches the fiducial
upper bound. A change in the assumed beaming geometry can change the
normalization and the inferred physical parameters. The correlated
analysis also retains substantial support for longer initial periods,
since fallback torques can alter the spin before the first XRT bin.

Neither calculation fits the individual variable excesses or excludes
contributions from other emission processes. The later afterglow remains
outside the fitted physical model. The 400--430 s portion can contain
overlapping components. The direct endpoint comparison and the
importance-reweighting diagnostic address this choice separately from
the temporal-block resampling; neither yields a unique decomposition.

\subsection{Conditional physical interpretation}
The fixed stellar parameters and the adopted disk torque define the scope
of the calculation. Away from the stellar-surface floor, the instantaneous
magnetospheric scaling is $r_{\rm m}\propto\xi B^{4/7}R_*^{12/7} M_*^{-1/7} \dot M_{\rm fb}^{-2/7}$. Thus, halving or doubling $\xi$ halves or doubles $r_{\rm m}$ at fixed
state; doubling the fallback normalization multiplies $r_{\rm m}$ by
$2^{-2/7}\simeq0.82$. In the disk-coupled branch,
$L_{\rm EM}\propto r_{\rm m}^{-2}$ at fixed $B,R_*,\Omega_*$, so the same
changes alter the instantaneous luminosity and can change the torque
regime. At fixed torques, $\dot\Omega_*\propto I_*^{-1}$. These scalings
explain the relevant sensitivity, but are not a substitute for refitting
the full evolution under changed stellar or disk parameters.

We quantify this dependence with one-at-a-time forward calculations
about the descriptive candidate, reported in Table~\ref{tab:sensitivity}.
The stored values of $P_0$, $B$, $t_{\rm fb}$ and $\epsilon_{\rm eff}$
are held fixed, and the spin is re-integrated from the same initial epoch
for each changed quantity. We vary $\xi$ over 0.5--2, the fallback
normalization by a factor of two, $I_*$ over
$(1$--$2)\times10^{45}\,\mathrm{g\,cm^2}$, $M_*$ over
$1.2$--$2M_\odot$, $R_*$ over 10--14 km and $\kappa_\chi$ over 1--2.
These checks isolate parameter dependence; the independently varied
$M_*,R_*,I_*$ values are not a sequence derived from a common equation
of state. No re-optimization or additional MCMC is performed.

The evolved fluxes show substantial sensitivity, including changes in
the timing of transitions between torque regimes. For example,
$\xi=2$ retains the ejector branch through 400 s, giving a flux 7.03
times the fiducial value, whereas the fiducial solution has entered the
propeller branch. For $I_*=2\times10^{45}\,\mathrm{g\,cm^2}$ the
400-s flux ratio is 9.23. These comparisons do not establish that the
altered curves fit the data or quantify parameter shifts after refitting.
They demonstrate why the reported intervals are conditional on the fixed
assumptions. Tightening the integration tolerance changes the reference
fluxes by less than $10^{-4}$ in relative terms.

The supplied fallback mass is not equated with accreted stellar mass.
Our constant-$M_*,R_*,I_*$ prescription does not test survival against
collapse. Such a test would require the retained mass and angular momentum,
and a consistent equation of state. Likewise, no timing-based radius
hierarchy or precession coherence condition is used to restrict the flux fit.}

{
\subsection{Comparison with fallback models of early X-ray bumps}
\citet{2024MNRAS.535.2482Y} report sample-distribution centres
$\log_{10}(t_1/\mathrm{s})=3.01$, $\log_{10}(B_0/\mathrm{G})=13.77$
and $P_0=1.04$ ms for their early-bump analysis.
They fit fallback emission together with prompt-tail and afterglow
components, adopting $R_*=10$ km and $\sin\chi=0.5$.
Their $\eta_{\rm mag}$ scales the fallback rate; it is not our
radiative/beaming factor $\epsilon_{\rm eff}$.
Our descriptive candidate instead gives $t_{\rm fb}=143$ source-frame
seconds, $B=3.04\times10^{15}$ G and $P_0=4.25$ ms; the correlated
posterior has medians 54.8 s, $1.56\times10^{15}$ G and 16.8 ms,
with a particularly broad period distribution.
Thus our estimates do not reproduce their typical field or spin.
Differences in fitted components, torque prescriptions, stellar assumptions
and time definitions preclude treating these numbers as a like-for-like
population test.

\section{Conclusions}
\label{sec:conclusions}

We describe a possible smooth magnetar contribution to the 
XRT observations in the observer-frame interval 80--430 s of GRB~241030A. The
fallback-modified spin evolution is converted consistently to the
unabsorbed 0.3--10 keV flux. Individual variable excesses are treated
statistically, and no periodicity or precession condition enters the
inference.

The correlated-residual posterior has median $P_0=16.8$ ms with central
68 per cent interval 2.76--286 ms; the corresponding medians of $B$,
$t_{\rm fb}$ and $\epsilon_{\rm eff}$ are $1.56\times10^{15}$ G,
54.8 source-frame seconds and 2.34. The sampler passes the stated
numerical diagnostics, but the broad initial-period distribution and
prior-sensitive residual timescale limit the physical interpretation.
Posterior predictive replication reveals discrepancies in the total
observed emission: replicated light curves typically have lower flux
levels, smaller overall dispersion and lower peaks than observed.
The posterior summaries therefore remain conditional on a residual
prescription whose adequacy is not supported by these checks; numerical
convergence does not validate the inferred physical decomposition.

A separate marginal descriptive estimate gives $P_0=4.25$ ms,
$B=3.04\times10^{15}$ G, $t_{\rm fb}=143$ source-frame seconds,
$M_{\rm norm}=8.99\times10^{-3}M_\odot$ and $\epsilon_{\rm eff}=81.2$,
at the fiducial efficiency ceiling. Its exploratory temporal-block
uncertainty is distinct from the posterior credible intervals. The
predictive assessment of the correlated model does not validate this
separate estimate or its bootstrap ranges. Its smooth flux is maximal
at the start of the fitted interval, 80 s, and declines thereafter;
the observed total-flux peak near 180 s is not identified as a
magnetar-component maximum. The fallback supply peaks near 198 s.
At 200 s the descriptive candidate is on the ejector branch and
enters a disk-coupled propeller phase near 255 s.

Joint posterior propagation gives estimated probabilities 0.552 and
0.834 for $r_{\rm m}<r_{\rm co}$ and $r_{\rm m}<r_{\rm LC}$ at 200 s.
The corresponding probabilities for each condition to hold at every
evaluated epoch in 80--430 s are 0.277 and 0.788. These probabilities
inherit the assumptions and limitations of the correlated-residual
posterior. The separate, finite-grid summaries do not establish
persistent positive disk torque. They provide no probability for a
timing-based radius hierarchy or coherent precession region.

These calculations illustrate a conditional descriptive interpretation
of a possible magnetar contribution, while exposing limitations in
its statistical separation from the variable emission. They do not
establish a unique emission decomposition, a precise birth spin, or
a precession origin of the variability. The assumed beaming geometry,
residual model, fixed stellar structure and possible afterglow overlap
remain relevant to the interpretation.}

\section*{Acknowledgements}

NF  acknowledges  financial  support  from UNAM-DGAPA-PAPIIT  through  grant  IN112525. The authors declare no conflicts of interest.

\section*{DATA AVAILABILITY}

The Swift/XRT and Swift/BAT data analysed in this work are publicly
available through the UK Swift Science Data Centre Burst Analyser for GRB~241030A at \url{https://www.swift.ac.uk/burst_analyser/01263718/}.

\bibliographystyle{mnras}
\bibliography{Bib_GRB160816A-GRB131108A}

\newpage

{
\begin{table}
\centering
\caption{Correlated-residual MCMC: independent uniform priors in the listed coordinates, posterior medians and central 68 per cent intervals from all 240000 cold production samples in 80--430 s. The $t_{\rm fb}$ and $\ell$ timescales are in the source and observer frames, respectively. The corner plot marginalizes over $M_{\rm norm},\tau,\ell,a$.}
\label{tab:fits}
\renewcommand{\arraystretch}{1.25}
\begin{tabular}{lcc}
\hline
Parameter & Prior & Median and 68 per cent interval \\
\hline
$\log_{10}(P_0/\mathrm{ms})$ & $[0,3]$ & $1.23^{+1.23}_{-0.79}$ \\
$\log_{10}(B/\mathrm{G})$ & $[13,16]$ & $15.19^{+0.41}_{-0.49}$ \\
$\log_{10}(t_{\rm fb}/\mathrm{s})$ & $[1,3.5]$ & $1.74^{+0.72}_{-0.29}$ \\
$\log_{10}(M_{\rm norm}/M_\odot)$ & $[-4,0]$ & $-0.95^{+0.66}_{-0.83}$ \\
$\log_{10}\epsilon_{\rm eff}$ & $[-3,\log_{10}(81.2)]$ & $0.37^{+0.95}_{-0.92}$ \\

%\cline{1-3}\cdashline
$\log_{10}(\tau/\mathrm{dex})$ & $[-2,1]$ & $-0.41^{+0.10}_{-0.09}$ \\
$\log_{10}(\ell/\mathrm{s})$ & $[-1,2.5]$ & $2.46^{+0.03}_{-0.06}$ \\
$\log_{10}[a/(1-a)]$ & $[-3,3]$ & $2.57^{+0.16}_{-0.15}$ \\
\hline\end{tabular}\end{table}

%\begin{table}
%\centering
%\caption{Marginalized posterior constraints obtained from modelling the XRT light curve of GRB 241030A within the magnetar scenario.}
%\label{tab:fits}
%\renewcommand{\arraystretch}{1.6}\addtolength{\tabcolsep}{3pt}
%\resizebox{0.3\columnwidth}{!}
%{
%\begin{tabular}{l c }
%\hline
%\hline
%Parameter & Value               \\ \hline \hline
%$\log_{10}\left(\frac{P_0}{\mathrm{s}}\right)$   & $-2.130^{+0.102}_{-0.085}$   \\
%$\log_{10}\left(\frac{B_{\rm }}{\mathrm{G}}\right)$ & $15.454^{+0.109}_{-0.086}$   \\
%$\log_{10}\left(\frac{t_{\rm fb}}{\mathrm{s}}\right)$ & $1.750^{+0.099}_{-0.097}$ \\
%$\log_{10}\epsilon_X$ & $-1.687^{+0.114}_{-0.087}$\\
%\hline
%\end{tabular}%
%}
%\end{table}

\begin{table}
\centering
\caption{Separate marginal descriptive estimate. The candidate column is the fit to the original data; the interval columns are empirical 16th--84th percentiles of 100 temporal-block refits each. They are not MCMC credible intervals. No boundary refits are removed. The effective efficiency is bounded above by 81.2.}
\label{tab:bootstrap}
\renewcommand{\arraystretch}{1.25}
\begin{tabular}{lrrrr}
\hline
Parameter & Candidate & 10 s blocks & 20 s blocks & 40 s blocks \\
\hline
$P_0$ (ms) & 4.25 & [3.88, 4.64] & [3.28, 4.37] & [1.00, 4.66] \\
$B/(10^{15}\,\mathrm{G})$ & 3.04 & [2.83, 3.39] & [2.61, 3.12] & [1.44, 3.27] \\
$t_{\rm fb}$ (s, source) & 143.1 & [118.6, 3029.5] & [103.2, 191.8] & [68.8, 448.8] \\
$M_{\rm norm}/M_\odot$ & 0.0090 & [0.0080, 0.4598] & [0.0081, 0.0870] & [0.0080, 0.1773] \\
$\epsilon_{\rm eff}$ & 81.20 & [80.88, 81.20] & [78.27, 81.20] & [9.10, 81.20] \\
$\tau$ (dex) & 0.342 & [0.264, 0.375] & [0.227, 0.375] & [0.165, 0.387] \\
\hline\end{tabular}\end{table}

\begin{figure}
{ \centering
%\resizebox*{0.65\textwidth}{0.55\textheight}
%{\includegraphics{GRB230307A.pdf}}\\
\resizebox*{0.8\textwidth}{0.8\textheight}
{\includegraphics{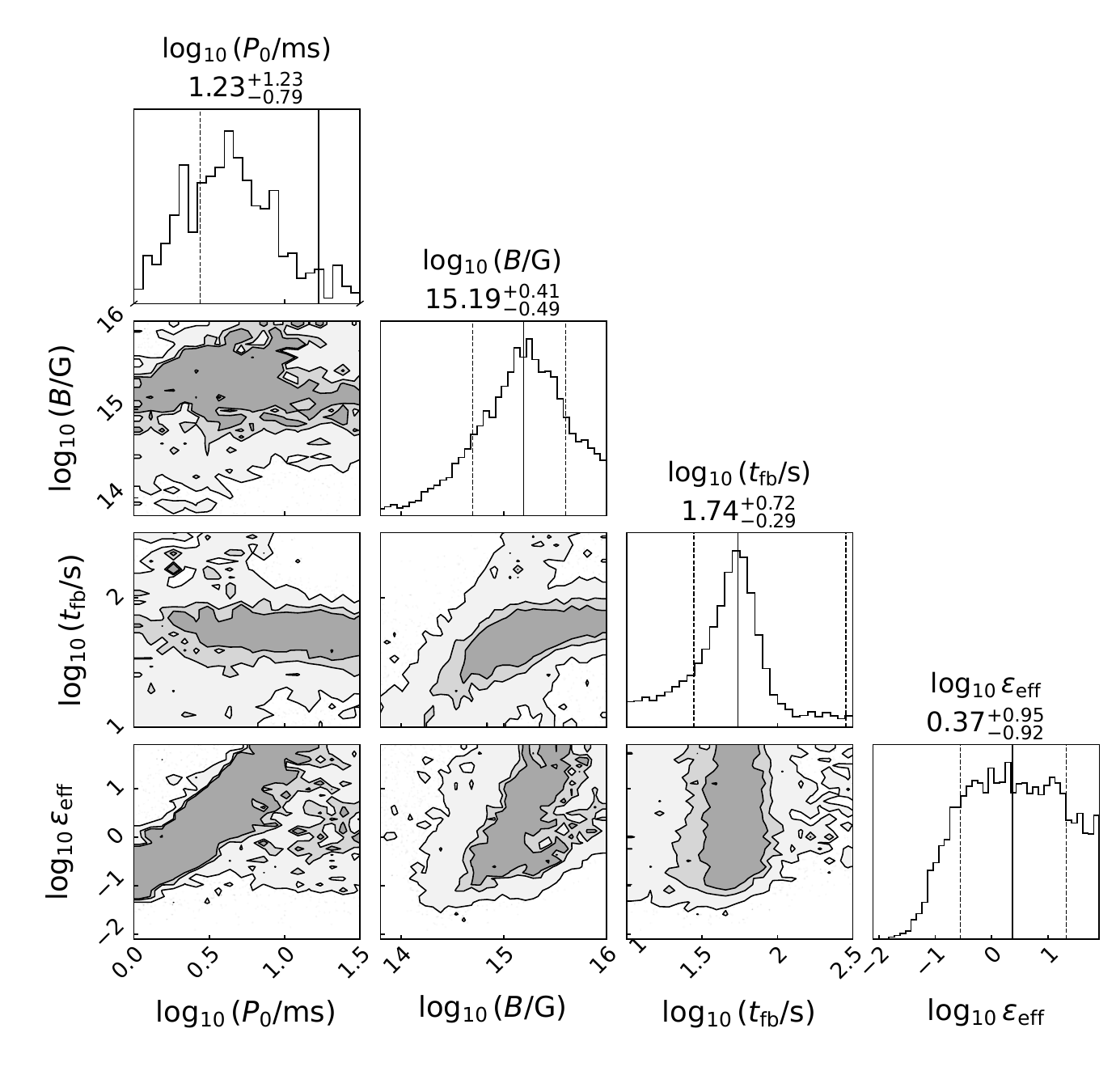}}
%\resizebox*{0.4\textwidth}{0.35\textheight}
%{\includegraphics{plot_magnetar_P.pdf}}
}
\caption{Marginal posterior distributions of four physical parameters for the correlated-residual magnetar model of GRB~241030A. The nested contours enclose approximately 50\%, 68\%, and 95\% of each two-dimensional marginal posterior in the displayed logarithmic coordinates, with darker shading toward higher density. One-dimensional histograms use 50 bins; two-dimensional histograms use 40 bins per axis over the original sampling bounds, without smoothing. Faint grey points show a uniformly thinned subset for visualization only. The diagonal panels show the one-dimensional distributions; black solid and dashed vertical lines indicate the median and the 16th and 84th percentiles, respectively. The values above the diagonal are medians and central 68\% intervals in these logarithmic coordinates. All 240,000 cold-chain production samples from four independent runs are pooled, after excluding warm-up. For readability, we display only the four parameters central to the physical discussion: $P_0$, $B$, $t_{\rm fb}$, and $\epsilon_{\rm eff}$. The fallback normalization $M_{\rm norm}$ and residual-model parameters $\tau$, $\ell$, and $a$ are sampled jointly and marginalized over in the displayed distributions; their posterior summaries are reported in Table~\ref{tab:fits}. The fallback timescale is in the source frame. Results are conditional on the adopted model and priors, including $\epsilon_{\rm eff}\leq81.2$. The displayed range of $\log_{10}(P_0/\mathrm{ms})$ is restricted to $[0,1.5]$, and that of $\log_{10}(t_{\rm fb}/\mathrm{s})$ to $[1,2.5]$; the displayed lower limit of $\log_{10}\epsilon_{\rm eff}$ is $-2.1$, and the displayed range of $\log_{10}(B/\mathrm{G})$ is $[13.8,16]$; all samples, including those outside this view, enter the histogram, contour thresholds, and reported quantiles. The lower ordinate limit of the $P_0$ marginal is nonzero (axis breaks), removing empty space below the curve without changing the density or samples. The full-range $P_0$ distribution, with its ordinate starting at zero, is shown in Figure~\ref{fig:p0-full}.}
 \label{fig1:mcmc}
\end{figure}

\begin{figure}
{ \centering
\resizebox*{0.85\textwidth}{0.5\textheight}
{\includegraphics{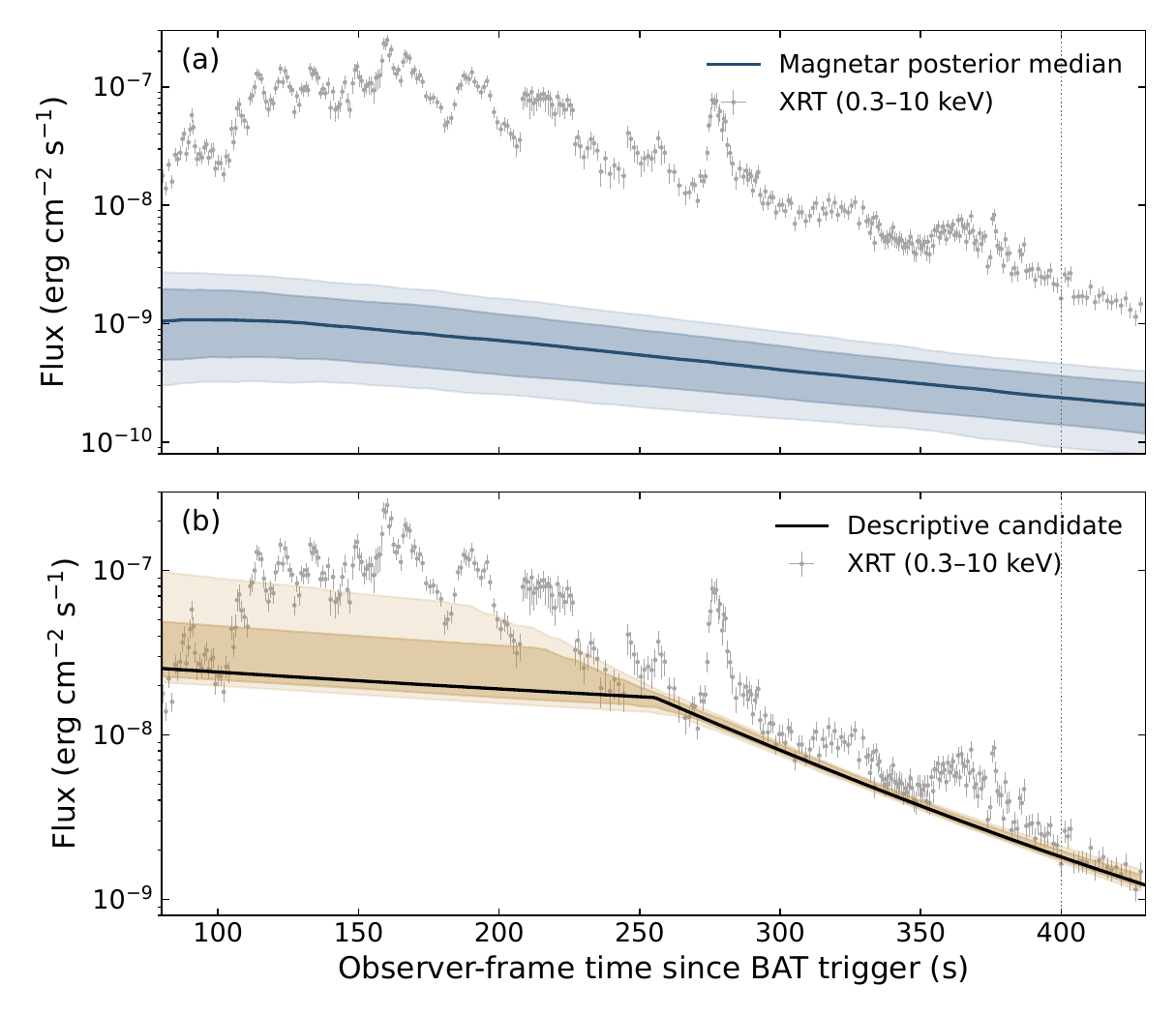}}\\
}
\caption{Two separate descriptions of the early XRT flux of GRB~241030A.
(a) Pointwise median of the smooth magnetar component from 1000 joint
draws of the correlated-residual posterior represented in Figure~1
and Table~1. The darker and lighter blue bands show the central
68 and 90 per cent pointwise credible intervals, respectively.
(b) Original-data marginal descriptive candidate, with darker and
lighter ochre bands showing the empirical 16th--84th and 5th--95th
percentiles, respectively, of 100 temporal-block refits with 20-s blocks.
These correspond to central 68 and 90 per cent bootstrap ranges,
not posterior credible intervals. The solid black curve is the
original-data candidate, not a bootstrap median.
Both panels show the original XRT flux uncertainties. The dotted
vertical line marks 400~s; both analyses use the 80--430~s selection.
The bands describe uncertainty in the magnetar component, not
predictive intervals for the total variable emission. No afterglow
model is subtracted or refitted.}
\label{fig:flux-current}
\end{figure}

\begin{figure}
{ \centering
\resizebox*{0.85\textwidth}{0.5\textheight}
{\includegraphics{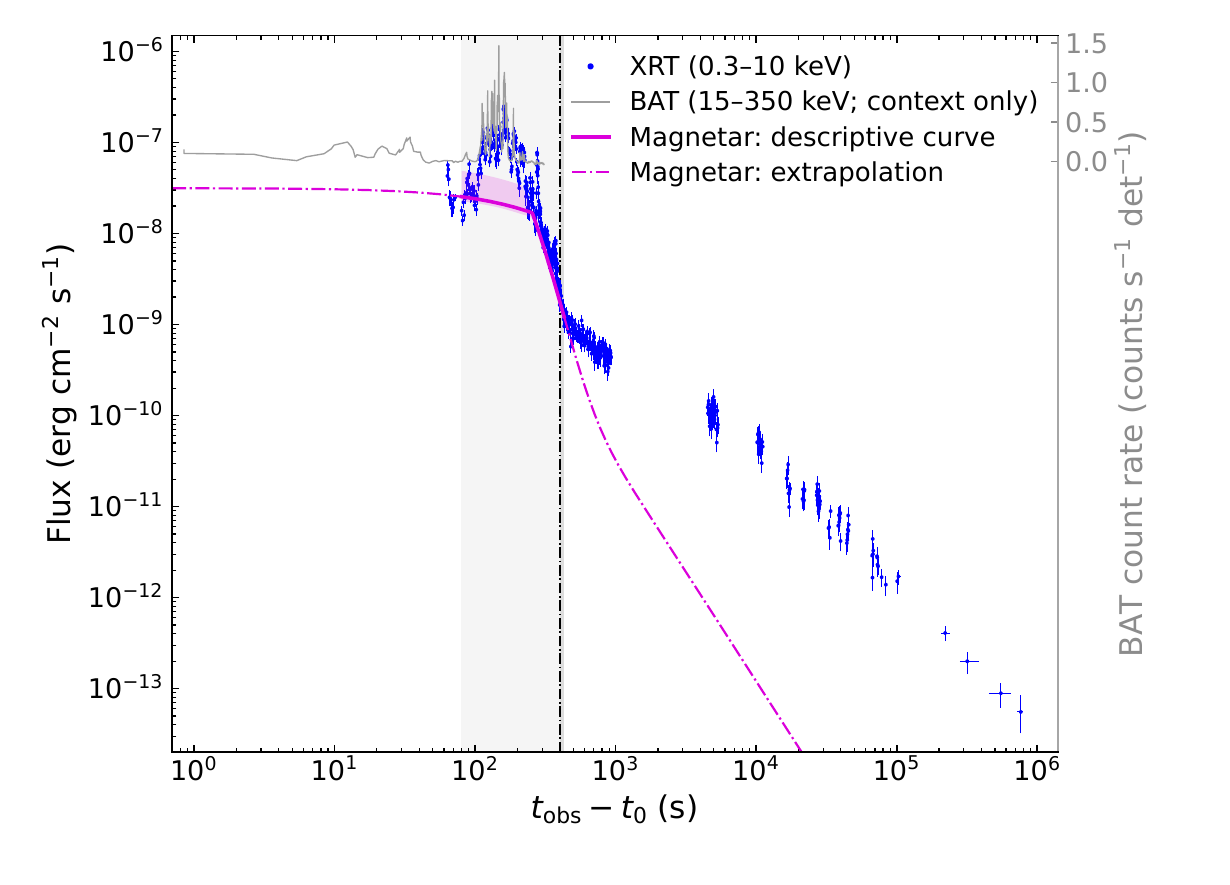}}\\
%\resizebox*{0.48\textwidth}{0.35\textheight}
%{\includegraphics{plot_magnetar_Rs_i_in.pdf}}
%\resizebox*{0.48\textwidth}{0.35\textheight}
%{\includegraphics{plot_magnetar_P.pdf}}
}
\caption{Swift/XRT unabsorbed 0.3--10 keV flux of GRB~241030A, shown
with the descriptive fallback-magnetar component (solid magenta curve)
and its extrapolation outside the fitted interval (dot--dashed magenta).
The magnetar curve uses the candidate parameters in Table~2 and is
identical to the descriptive curve in Figure~\ref{fig:flux-current}(b). The magenta band
shows the pointwise empirical 16th--84th percentiles of the existing
100 refits with 20-s temporal blocks; it is not an MCMC credible band.
The grey shading marks the 80--430 s fitting interval, with darker
shading over 400--430 s, and the black dot--dashed line marks 400 s.
The Swift/BAT 15--350 keV count rate is displayed on the right-hand
axis for temporal context only and is not included in the fit.
The magnetar curve represents observer-frame flux,
$F_{\rm mag}=\epsilon_{\rm eff}L_{\rm EM}/(4\pi D_L^2)$, with
$F_{\rm mag}(200\,\mathrm{s})=1.902\times10^{-8}$ erg cm$^{-2}$ s$^{-1}$.
No afterglow component is fitted or displayed.}
 \label{fig2:radii}
\end{figure}

\begin{figure}
{ \centering
\resizebox*{0.85\textwidth}{0.5\textheight}
{\includegraphics{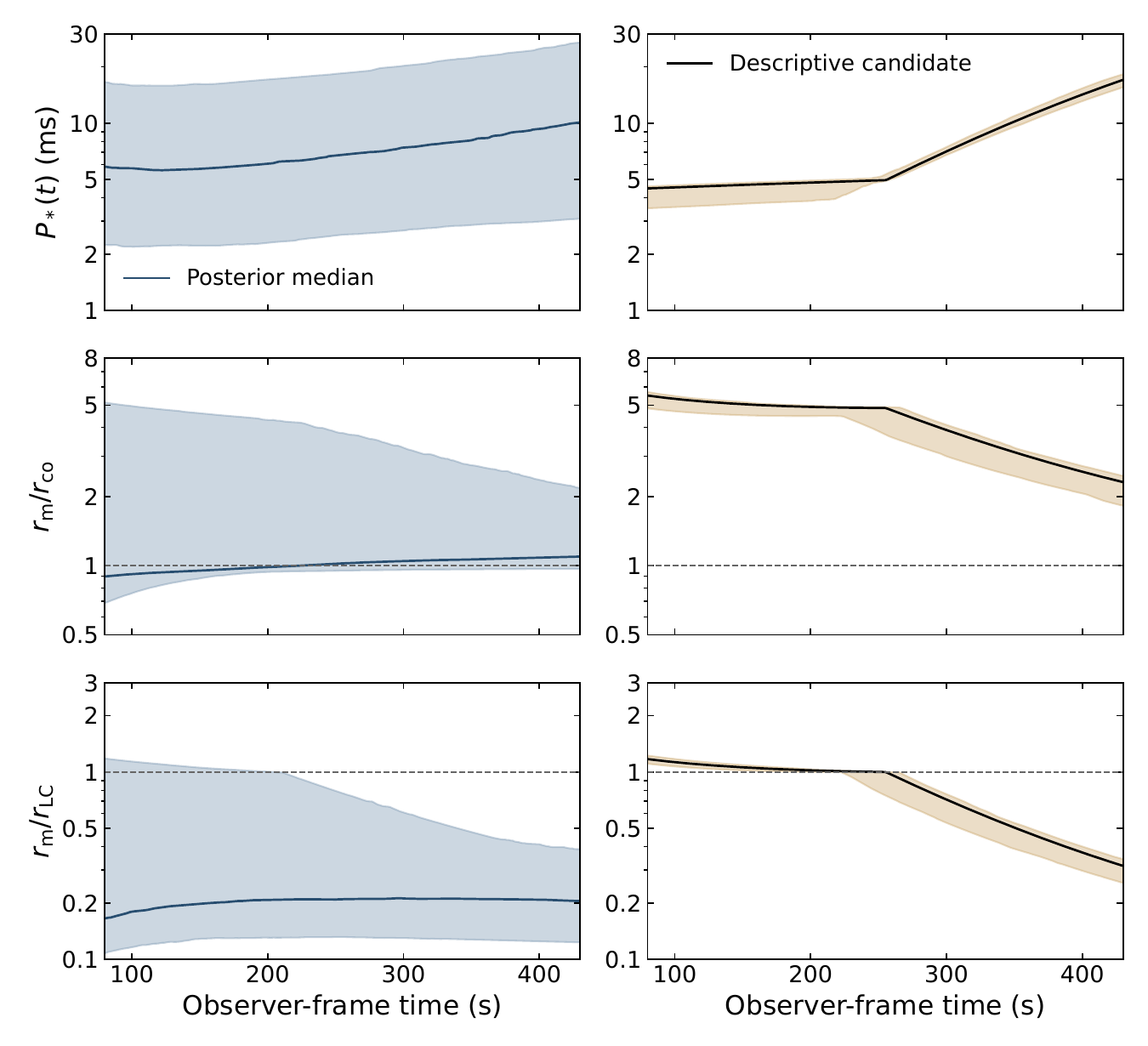}}\\
}
\caption{Evolution of the stellar spin period $P_*(t)$ (top), the
magnetospheric-to-corotation radius ratio $r_{\rm m}/r_{\rm co}$
(middle), and the magnetospheric-to-light-cylinder radius ratio
$r_{\rm m}/r_{\rm LC}$ (bottom), as functions of observer-frame time.
Left: the blue curves and shaded bands show the pointwise medians
and central 68 per cent credible intervals obtained by propagating
1000 joint draws from the correlated-residual posterior.
Right: the black curves show the descriptive candidate, while the
ochre bands span the pointwise 16th--84th percentiles of the 100
temporal-block refits with 20-s blocks. These empirical bootstrap
ranges are distinct from the posterior credible intervals.
Horizontal dashed lines mark unity in the radius ratios.
Positive disk torque requires both $r_{\rm m}<r_{\rm co}$ and
$r_{\rm m}<r_{\rm LC}$ in the adopted prescription.
Corresponding panels use identical logarithmic vertical scales.}
\label{fig:dynamics-current}
\end{figure}

\begin{table}
\centering
\caption{Sensitivity of the marginal descriptive estimate to the upper
observer-frame time limit. Both windows use the same current physical
model, six-parameter marginal objective, bounds, and original bins.
These are optimized point estimates without new endpoint-specific
uncertainty intervals; the correlated-posterior diagnostic is discussed
separately in the text.}
\label{tab:endpoint}
\begin{tabular}{lcc}
\hline
Quantity & 80--430 s & 80--400 s \\
\hline
Number of bins & 380 & 361 \\
$P_0$ (ms) & 4.25 & 4.10 \\
$B$ ($10^{15}$ G) & 3.04 & 2.79 \\
$t_{\rm fb}$ (source-frame s) & 143.1 & 122.7 \\
$M_{\rm norm}$ ($10^{-3}M_\odot$) & 8.99 & 8.07 \\
$\tau$ (dex) & 0.342 & 0.351 \\
$\epsilon_{\rm eff}$ & 81.2 & 81.2 \\
\hline
\end{tabular}
\end{table}

\begin{table*}
\centering
\caption{One-at-a-time forward sensitivity about the descriptive candidate.
Times are in the observer frame. $F_0(t)$ is the fiducial descriptive flux,
$1.902\times10^{-8}$ and $1.810\times10^{-9}$ erg cm$^{-2}$ s$^{-1}$
at 200 and 400 s, respectively. Only the listed quantity changes;
$M_{{\rm norm},0}=8.99\times10^{-3}M_\odot$. The spin evolution is
re-integrated, while the remaining candidate parameters are held fixed.
These are not refits or posterior intervals, and independent changes of
stellar mass, radius and inertia do not define an equation-of-state sequence.}
\label{tab:sensitivity}
\begin{tabular}{lrrrr}
\hline
Change & $P_*(200)$ & $F(200)/F_0(200)$ & $P_*(400)$ & $F(400)/F_0(400)$ \\
 & (ms) & & (ms) & \\
\hline
Fiducial & 4.82 & 1.00 & 14.22 & 1.00 \\
$\xi=0.5$ & 7.97 & 1.40 & 14.28 & 3.97 \\
$\xi=2$ & 4.82 & 1.00 & 5.33 & 7.03 \\
$M_{\rm norm}=0.5M_{{\rm norm},0}$ & 4.82 & 1.00 & 5.33 & 7.03 \\
$M_{\rm norm}=2M_{{\rm norm},0}$ & 15.65 & 0.13 & 37.70 & 0.21 \\
$I_*=1.0\times10^{45}\,\mathrm{g\,cm^2}$ & 5.78 & 0.66 & 30.15 & 0.22 \\
$I_*=2.0\times10^{45}\,\mathrm{g\,cm^2}$ & 4.62 & 1.18 & 4.97 & 9.23 \\
$M_*=1.2M_\odot$ & 4.82 & 1.00 & 5.33 & 7.03 \\
$M_*=2.0M_\odot$ & 6.37 & 0.61 & 20.33 & 0.54 \\
$R_*=10\,\mathrm{km}$ & 8.06 & 0.21 & 15.40 & 0.53 \\
$R_*=14\,\mathrm{km}$ & 5.57 & 1.41 & 6.64 & 7.33 \\
$\kappa_\chi=2$ & 8.72 & 0.58 & 31.54 & 0.41 \\
\hline
\end{tabular}
\end{table*}

\clearpage
\appendix
\section{Posterior predictive assessment}
\label{app:ppc}

This appendix documents the posterior predictive calculation reported
in Section~\ref{sec:correlated-posterior}. We use the same 1000 joint
posterior draws propagated for the dynamics, with 250 from each of the
four independent cold runs. For each draw, we generate
$\boldsymbol z^{\rm rep}\sim\mathcal N(0,R)$ at the original bin centres
and set
\begin{equation}
F_i^{\rm rep}=F_{{\rm mag},i}\,
10^{H_i^{-1}[\Phi(z_i^{\rm rep})]},
\label{eq:ppc-replication}
\end{equation}
where $F_{{\rm mag},i}$ is the bin-averaged physical flux and $H_i$ is
the exGaussian residual cumulative distribution defined in
Section~\ref{sec:model-fitting}. The original logarithmic error estimates
are held fixed and already enter $H_i$; no additional measurement-noise
realization is added. The Gaussian vector is drawn from the stationary
copula without conditioning on the observed residual path. No refitting
is performed.

With $y_i=\log_{10}[F_i/(\mathrm{erg\,cm^{-2}\,s^{-1}})]$, the three
pre-specified summaries are the sample standard deviation of $y_i$,
its maximum, and
\begin{equation}
T_{10-40}=\left[\frac{1}{|\mathcal P|}
\sum_{(i,j)\in\mathcal P}(y_j-y_i)^2\right]^{1/2},
\qquad
\mathcal P=\{(i,j):10\,\mathrm{s}\leq
t_{{\rm obs},j}-t_{{\rm obs},i}\leq40\,\mathrm{s}\}.
\label{eq:ppc-temporal}
\end{equation}
All 11907 pairs have equal weight. Table~\ref{tab:ppc} gives the
observed and replicated summaries; Figure~\ref{fig:ppc} shows the
pointwise flux predictions and the summary distributions.
The upper-tail fraction $p_B=N(T^{\rm rep}\geq T^{\rm obs})/1000$
is not a frequentist-calibrated significance level.

\begin{table}
\centering
\caption{Posterior predictive summaries from 1000 replicated observed
light curves. The replicate range comprises the 5th--95th percentiles.
All summaries use logarithmic flux. The column $p_B$ gives the
upper-tail fraction defined in the text.}
\label{tab:ppc}
\begin{tabular}{lrrrr}
\hline
Statistic & Observed & Replicate median & Replicate range & $p_B$\\
\hline
Standard deviation (dex) & 0.593 & 0.277 & [0.114, 0.587] & 0.048\\
Maximum $y_i$ & $-6.603$ & $-8.382$ & [$-9.148$, $-6.946$] & 0.032\\
$T_{10-40}$ (dex) & 0.310 & 0.149 & [0.068, 0.349] & 0.071\\
\hline
\end{tabular}
\end{table}

Only 79 of 380 measured fluxes lie within the central 90 per cent
pointwise predictive intervals, while 301 lie above them. This is a
descriptive in-sample comparison, not an independent-bin coverage test.
The shortcomings documented here limit the physical interpretation of
the correlated posterior. The calculation does not assess or validate
the separate descriptive estimate and its temporal-block bootstrap ranges.

\begin{figure}
\centering
\includegraphics[width=0.7\linewidth, height=0.35\textheight]{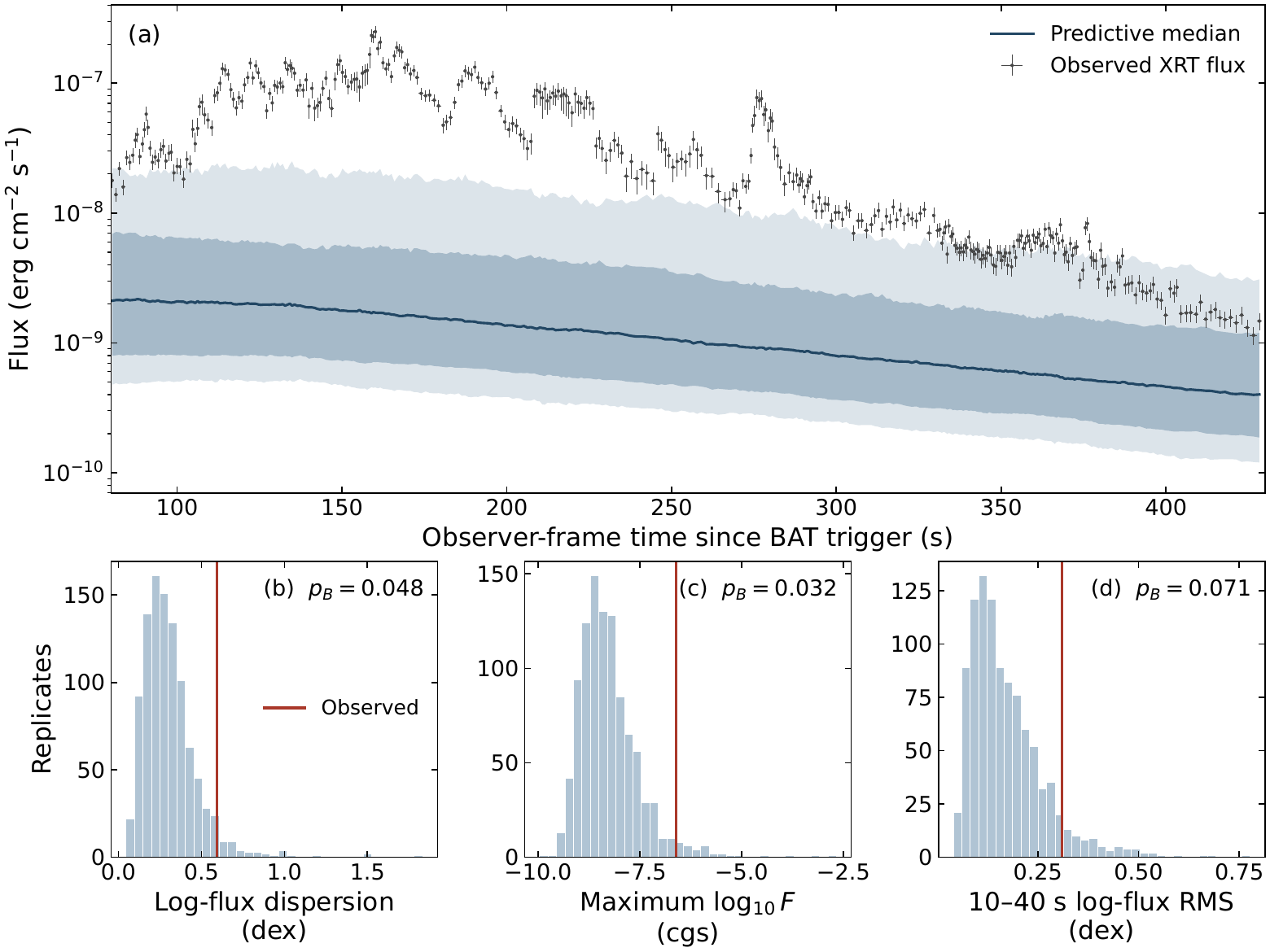}
\caption{Posterior predictive assessment of the correlated-residual
description of GRB~241030A. (a) Observed XRT flux and central 68 and
90 per cent pointwise predictive intervals (darker and lighter blue,
respectively) for replicated observed flux, including the statistical
excess and measured log-flux error
approximation. The blue line is the predictive median, not the
magnetar-component median. (b--d) Distributions of the three
pre-specified summaries across 1000 replications; red vertical
lines indicate the observed values. All replicated summary values
enter the displayed histograms. The upper-tail fractions $p_B$ are
defined in the text. These checks use the existing joint posterior
without refitting and do not assess the separate descriptive estimator.}
\label{fig:ppc}
\end{figure}

\clearpage
\newpage

\section{Full initial-period distribution}
\label{sec:full-p0}
Figure~\ref{fig:p0-full} displays the complete sampled initial-period
range, complementing the restricted view in Figure~\ref{fig1:mcmc}.
It uses all 240000 retained cold-chain production samples, with
unchanged weights and no filtering. The median and central 68 per cent
interval are $P_0=16.8$ ms and 2.76--286 ms. Approximately 28.9 per cent
of the samples have $P_0>100$ ms, while 42.7 per cent lie above the
31.6-ms upper display limit in Figure~\ref{fig1:mcmc}. The extended
view changes neither the posterior nor its predictive limitations.

\begin{figure*}
\centering
\includegraphics[width=0.7\linewidth, height=0.35\textheight]{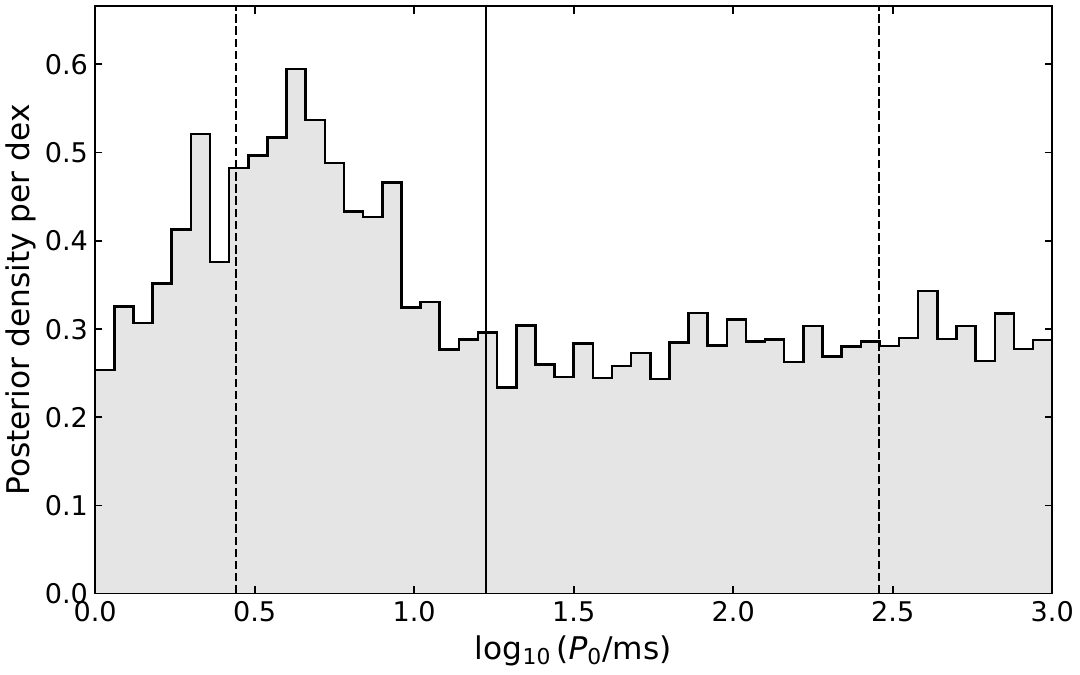}
\caption{Full marginal distribution of $\log_{10}(P_0/\mathrm{ms})$
from the same four independent cold runs used in Table~\ref{tab:fits}
and Figure~\ref{fig1:mcmc}. The histogram uses 50 equal-width bins
across the complete prior domain $0\leq\log_{10}(P_0/\mathrm{ms})\leq3$,
corresponding to $1\leq P_0\leq1000$ ms, without smoothing or
sample removal. The ordinate is probability density per dex and
starts at zero. The solid vertical line marks the median, and the
dashed lines mark the 16th and 84th percentiles. All physical and
residual parameters other than $P_0$ are marginalized over.
The distribution remains conditional on the adopted model and
priors; displaying its full extent does not resolve the posterior
predictive discrepancies documented in Appendix~\ref{app:ppc}.}
\label{fig:p0-full}
\end{figure*}

}

\end{document}